\documentclass[fleqn,usenatbib]{mnras}

\usepackage{newtxtext,newtxmath}

\usepackage[T1]{fontenc}
\usepackage{float}
\usepackage{placeins}
\usepackage{caption}
\usepackage{subfigure}
\usepackage{subcaption}
\usepackage{pdflscape}
\DeclareRobustCommand{\VAN}[3]{#2}
\let\VANthebibliography\thebibliography
\def\thebibliography{\DeclareRobustCommand{\VAN}[3]{##3}\VANthebibliography}
\usepackage{graphicx}	
\usepackage{amsmath}	

\title[Broken Expectations]{Broken Expectations: The Effects of Modelling Assumptions on the Inferred Dark Matter Distribution in the Milky Way's Satellites}

\author[K. Tchiorniy and A. Genina]{
Kristian Tchiorniy$^{1,2}$\thanks{E-mail: kristian.tchiorniy@tum.de}
and Anna Genina$^{1,3}$\thanks{E-mail: agenina@ed.ac.uk}
\\
$^{1}$Max Planck Institute for Astrophysics, Karl-Schwarzschild-Str. 1, 85748, Garching\\
$^{2}$Physics Department, Technical University of Munich, James-Franck-Str. 1, 85748, Garching\\
$^{3}$Institute for Astronomy, University of Edinburgh, Royal Observatory, Blackford Hill, Edinburgh EH9 3HJ, UK \\
}

\date{Accepted XXX. Received YYY; in original form ZZZ}

\pubyear{\the\year{}}

\begin{document}
\label{firstpage}
\pagerange{\pageref{firstpage}--\pageref{lastpage}}
\maketitle

\begin{abstract}

The spherical Jeans equation is commonly used to infer dark matter distributions in dwarf spheroidal satellites of the Milky Way to constrain the nature of dark matter. One of its assumptions is that of dynamical equilibrium while the dwarfs are under the influence of Galactic tides. We carry out tailored simulations of Carina, Draco, Fornax, Sculptor and Ursa Minor and test the accuracy of dark matter density profiles and annihilation rates (J-factors) recovered with the Jeans analysis code {\sc pyGravSphere}. We find that tides do not significantly affect the quality of density profile inference; however, {\sc pyGravSphere} tends to underestimate the inner densities of dwarf galaxies, which, together with tidal mass loss, leads to an inference of flatter density slopes, although all of our dwarfs have cuspy Navarro-Frenk-White haloes. This is because the default broken power-law model is unable to describe the outer halo density profile. The recovered J-factors are generally underestimated. While the difference with the true J-factor is small, the error bars are also often underestimated. We also test the accuracy of the Wolf et al. 2010 mass estimator and find that it can be sensitive to orbital stage and eccentricity. Still, for our sample of dwarf galaxies, the estimates agree with the truth within 10~per~cent. Consistency of our simulated dwarfs with the mass-concentration relation in $\Lambda$CDM requires a light Milky Way, or limited action of tides, which may be in tension with a ``tidal stirring'' origin of dwarf spheroidals.

\end{abstract}




\section{Introduction}
Within the standard $\Lambda$-Cold Dark Matter ($\Lambda$CDM) model of cosmology \citep{fof}, dark matter makes up $\sim 75$~per~cent of the matter density in our Universe, yet its nature remains unknown \citep{planck}. In the last two decades, dwarf galaxies have become of great interest in dark matter searches (\citealt{baltz_jfac, evans_jfac,jfactors_hooper,sanchez_conde_jfac,strigari_jfactors}; see \citealt{strigari_review} for a review). Their high dynamical mass-to-light ratios \citep{mateo} imply that the internal kinematics of these objects are strongly dark matter dominated \citep{battaglia_helmi_review,walker_review,battaglia_review}, making it possible to study the nature of dark matter phenomenologically through its dynamical effect on the baryonic matter in dwarf galaxies \citep{wilkinson_models,lokas2002}. Moreover, since the Milky Way is currently known to host $\gtrsim$60 confirmed and candidate satellite dwarf galaxies \citep{bechtol_des,drlica-wagner,subaru_candidates,delve, pace_local_volume}, their relative proximity implies high fluxes of the by-products of dark matter annihilation or decay, making it possible to constrain the particle physics nature of dark matter with gamma and X-ray instruments such as Fermi \citep{fermi,fermi_dwarf}, the Cherenkov Telescope Array \citep{lefranc_cta,saturni_cta}, XMM-Newton and eRosita \citep{xmm1,xmm2,ando_xray}.

A key property relevant for dark matter studies is its density distribution near the centres of galaxies. In particular, some $N$-body dark-matter-only \citep{nfw96,aquarius} and galaxy formation simulations in $\Lambda$CDM \citep{noproblem,sownak_cores} predict that the dark matter density should be increasing towards the halo centres, forming a ``cusp'' ($\rho \propto r^{-1}$). Other simulations\footnote{See \citet{alejandro_cores} on the likely source of difference between simulation predictions.} \citep{nihao_cores,firecores, orkney_cores,jackson_horizon}, as well as some measurements of rotation curves in nearby dwarf galaxies, seem to suggest that the dark matter density is constant at galaxy centres \citep{floresprimack,Moorecore, oh2015,adamsrotcurv}, forming a ``core'' ($\rho \propto r^{0}$). This has become known as the ``core-cusp problem'' \citep{corecusp}, which remains a challenge to the standard model of cosmology (see \citet{sales_wetzel_fattahi} for a review). Two main solutions have been proposed to this problem: one is that bursty supernovae feedback in shallow dwarf galaxy potentials is responsible for lowering the dark matter density over time \citep{navarro_eke_frenk, read_gilmore, mashchenko, pontzen_governato, brooks_zolotov, read_agertz_collins}, the other is that dark matter is not cold, or can interact non-gravitationally, producing a core at galaxy centres \citep{spergel_steinhardt}. The latter explanation has motivated many studies in alternative dark matter models \citep{maccio_wdm, sidmcores, schive, chan_may_fuzzy}. The shortcoming of some of these models, however, is that they tend to produce dark matter cores along the full range of dark matter halo masses and in all halos, whereas latest rotation curve measurements imply a variety of dark matter distribution -- cusped, cored and something in-between \citep{de2010case, santos_diversity}. Some models are additionally inconsistent with sizes and density distributions inferred for nearby dwarfs \citep{correa2021constraining, dalal2022excluding, fuzzy_heavy}. As such, the ``core-cusp problem'' in recent years has transformed into a ``rotation curve diversity problem'' \citep{oman,relatores}, requiring a solution that can produce a variety of dark matter density distributions, with self-interacting dark matter, featuring core formation and gravothermal core collapse that can be accelerated by the presence of tidal forces \citep{Kahlhoefer_2019,nishikawa2020accelerated}, becoming a popular candidate \citep{kamada,tulin2018dark,roberts2024gravothermal}.

One caveat to the diversity problem is that shallow or cored density distributions are inferred through rotation curves, which are prone to a number of systematic effects that tend to bias the measurement towards cores \citep{fatal_attraction, oman_bias, roper}. A different way to derive the dark matter mass distribution is through the modelling of stellar kinematics \citep{battaglia_review}. Such inferences typically involve studies of dwarf spheroidal satellites of the Milky Way due to their proximity, which allows to resolve individual stars and to obtain high-resolution spectroscopic velocities and photometric profiles required in the modelling. To date, there has not been a definitive detection of a dark matter core in a dwarf galaxy through stellar kinematics \citep{strigari_core, Breddels_2013, genina}, though Sculptor and Fornax have been identified as likely candidates for hosting a dark matter core \citep{goerdt_core_fornax, battagliasculptor, wp11,amoriscoevans,agnelloevans, Jardel_2012,amoriscoFornaxcore, breddels_sculptor,pascale19_actionbased, schwarz_fornax}, which includes Jeans models, Schwarzschild models \citep{schwarz} and distribution function-based techniques. Using the non-parametric Jeans modelling technique {\sc GravSphere} \citep{Read_2017, Genina_2020, binulator}, \citet{Read2_2019} have shown that from a sample of dwarf spheroidal galaxies of the Milky Way, only Fornax has a dark matter density that could be compatible with a dark matter core, or a shallow cusp. The relatively large stellar mass of Fornax compared to other dwarf spheroidals also supports the hypothesis that the low dark matter density in Fornax is the result of stellar feedback \citep{dicintio_cores}.

The difficulty in obtaining a convincing detection of a dark matter core through the modelling of stellar kinematics is three-fold. Firstly, the kinematic data are typically limited to the line-of-sight velocities, leading to a so-called mass-anisotropy degeneracy in the popular Jeans dynamical models, meaning that the uncertainty on the transverse motions of the stars translates to the uncertainty on the enclosed mass profile. Even with the proper motion data becoming available with Gaia \citep{gaia_edr3, gaia_edr3_pm}, most dwarfs of interest are too far away to allow for sufficiently small proper motion errors \citep{StrigariFrenkWhite_proper}; however, some works have combined the Gaia proper motions with the Hubble Space Telescope data to reduce these errors \citep{massari_draco}, while some utilised the longer-baseline HST data, reaching deeper magnitude limits than Gaia \citep{vitral_draco, vitral_sculptor}. Secondly, the kinematic and photometric samples of stars in the Milky Way dwarf spheroidals are typically small, owing partially to the already small stellar masses of these galaxies (typically < 10$^7 M_{\odot}$). \citet{chang_necib} and \citet{guerra_strigari} have shown that samples of $\sim 10^5$ kinematic tracers are required for a convincing detection of a dark matter core\footnote{Although we note that these studies consider only a single available tracer population and only use the constraints of the second moment of the velocity distribution.}, although this may depend on core size.  Finally, dynamical models (particularly those based on the Jeans equation) typically make a number of assumptions which are known to, or might to some extent, be broken in Milky Way's dwarf satellites, such as spherical symmetry \citep{McConnachie_2012} and dynamical equilibrium \citep{newberg_sag, fornax_merger}. This introduces a degree of scepticism in the interpretation of the results of dynamical models which do not account for these effects and has led several works to evaluate how cored or cuspy dark matter haloes are based on their densities below the half-light radius compared to $\Lambda$CDM predictions -- a quantity that can be evaluated more reliably than the innermost log-slope of the dark matter density distribution \citep{Read2_2019,gregory_tucana,binulator,charles,deleo}. Although rotation can be accounted for in Jeans and other types of models (e.g. \citealt{Zhu_2016}), its effects when modelling Milky Way's dwarf spheroidals are commonly assumed to be negligible, with observational studies confirming that rotation is typically insignificant compared to the velocity dispersion \citep{wheeler_rotation, dwarf_rotation}. The parameterization of the dynamical model and modelling technique can also lead to significant differences in the inferred dark matter distribution \citep{breaking_beta}. 

It is now well understood that asphericity in the stellar distribution can bias the results of spherical Jeans analyses \citep{kowal_nonsph}. It has been demonstrated both in the full profile recovery and in dynamical mass estimators at a given radius \citep{lokas_tides, laporte_wp, campbell, genina, Genina_2020} that the recovered mass profiles are often biased high when a dwarf is viewed along its major axis and low when viewed along the minor axis; however, the 3-D shapes of dwarf satellites of the Milky Way are not known due to the lack of availability of reliable distance measurements to individual stars. For most spectroscopic samples of dwarf spheroidals (< 1000 stars), the uncertainty associated with sparse data samples likely encompasses the modelling bias due to the dwarfs' asphericity. Fornax, however, has a sample size of $\approx$2500 stars \citep{fornax_stellar}, such that bias associated with asphericity may be more statistically significant. Some information can be gained from the projected galaxy shape and \citet{hayashi}, using axisymmetric Jeans modelling, obtain a central density for Fornax that has a factor of 2 larger $2\sigma$ error bars than that in the spherical Jeans modelling of \citet{Read2_2019}. This is due to significantly more freedom allowed by their models, which are also unconstrained by the higher moments of the distribution function. The result of \citet{hayashi} makes the inner dark matter density of Fornax compatible with an NFW cusp within 2$\sigma$. 

The uncertainty in recovered density profiles also translates to the uncertainty in J-factors -- a quantity that scales as the dark matter density squared and represents the expected flux of the by-products of dark matter annihilation that can be directly searched for with gamma-ray instruments \citep{conrad}. Spherical Jeans analysis has in the past been employed to estimate the J-factors of Milky Way's satellite galaxies. In particular, \citet{fermi_dwarf} combine the J-factors obtained through spherical Jeans modelling by \citet{pace_strigari} with scaling relation-derived J-factors for a sample of ultra-faint dwarfs to find a weak tension with WIMP interpretation of the Galactic Centre excess. Putting such a constraint would not be possible if J-factor uncertainties are larger than those suggested by the spherical Jeans modelling \citep{sanders_nonsph_j} or if the inferred densities are biased high. This makes understanding the limitations of the commonly used spherical Jeans models a particularly pressing issue. 

A largely unexplored source of uncertainty in the literature is the effect of Milky Way's tidal forces on the inferred dark matter distribution in the inner regions of dwarf galaxies, particularly for several satellites in a consistent Milky Way and cosmological framework; however, for individual satellites, several works have found limited impact by tides and accurate mass profile recovery given effective removal of unbound stars \citep{klimentowski, klimentowski2, serra_interloper, disequilibrium_ural, battagliaFornaxtides, Iorio_2019,  deleo}. Cosmological galaxy formation simulations in $\Lambda$CDM have made it possible to test the accuracy of modern dynamical models on simulated dwarfs. \citet{Genina_2020} have tested the {\sc pyGravSphere} method on Fornax-like dwarfs in the APOSTLE \citep{sawalapuzzles, fattahi_apostle} simulations (A Project for Simulating the Local Environment) and found no significant effect on the inner profiles and a significant effect on the outermost profiles, primarily caused by the limited priors to model the steepening of dark matter slope in the outer halo caused by tides \citep{penarrubiatides}. \citet{wenting} have tested Jeans anisotropic modelling \citep{jam} on Sagittarius-like dwarfs with the AURIGA simulations \citep{auriga} and found that their enclosed masses and inner density profiles were typically underestimated due to contraction motion caused by tidal effects. \citet{nguyen_fire_tides} tested a Graph Neural Posterior Estimator on a sample of dwarfs from the FIRE-2 Latte simulations \citep{latte_wetzel} and have identified trends of the method's effectiveness with the orbital pericentres, how many pericentres have been completed and the current position with respect to the host galaxy, with stronger tidal effects leading to an overestimation of the dwarf galaxies' masses and maximum circular velocities. Nevertheless, it is unclear whether these results are applicable to Milky Way's dwarfs, as it is unlikely that any given simulation would reproduce both the observed properties of the dwarf galaxies and their present-day orbits, thus making them unconvincing analogues of Milky Way's satellites. 

As in the past works by \citet{battagliaFornaxtides} and \citet{Iorio_2019}, who studied tidal effects on Fornax and Sculptor dSphs, we circumvent these issues by producing a suite of tailored simulations of Carina, Draco, Fornax, Sculptor and Ursa Minor in two disky Milky Way potentials from \citet{Bovy_2015}. Our simulated dwarf galaxies follow orbits in these potentials as derived from Gaia EDR3 proper motions \citep{battaglia_edr3} and, at $z=0$, have stellar masses, half-light radii and line-of-sight velocity dispersions consistent with current observations. Our dwarfs are all set up to have spherical  Navarro-Frenk-White (NFW) dark matter density profiles initially \citep{nfw96} and stellar components following a Plummer distribution \citep{plummer}. The dark matter content of the dwarfs is resolved with $10^7$ particles to avoid the effects of artificial disruption \citep{van_den_Bosch_2018}. In future work, we will aim to replicate this experiment with cored dark matter density profiles. For now, we aim to establish whether current observations of Milky Way's classical dwarfs can be replicated in a $\Lambda$CDM cosmology where cores \textit{do not} form. We then focus exclusively on the effect of Milky Way's tidal forces on the ability of spherical Jeans models to recover underlying dark matter density distributions of dwarf satellites and the accuracy of their recovered J-factors. We do not consider the effects of unbound stars in contaminating the kinematic sample and focus instead on the deformation of the dwarf spheroidal system due to tides and the tidal heating of particle orbits that need not necessarily leave the stars unbound. We make our sample of simulated dwarf spheroidals publicly available for tests of alternative dynamical models with varying kinematic sample selections.

This work is structured as follows: in Section~\ref{methods}, we describe the setup of our dwarf galaxy simulations (\ref{simulations_section}), the {\sc pyGravSphere} spherical Jeans modelling technique (\ref{gravspheresection}), the \citet{Wolf_2010} mass estimators (\ref{wolfsection}) and the calculation of the J-factors (\ref{sec:jfac_theory}). In Section~\ref{results_section}, we show the accuracy of the \citet{Wolf_2010} mass estimator as a function of Milky Way mass and orbital stage (\ref{sec:wolf}) and present the density profiles recovered with {\sc pyGravSphere} (\ref{sec:pyg_results}). The comparison between the inferred inner densities of our simulated dwarfs with the results of \citet{Read2_2019} is presented in Section~\ref{sec_densities} and the accuracy of the J-factors is evaluated in Section~\ref{section_jfactors}. In Section~\ref{conclusions_section}, we summarise our findings and conclude.

\section{Methods}
\label{methods}
\subsection{Tailored dwarf spheroidal simulations}
\label{simulations_section}
For our study of dSphs orbiting the Milky Way, we consider three key components vital for the simulations we would like to carry out: the Milky Way potential, the initial conditions of the dwarf galaxies and the numerical parameters of the simulations. We discuss these in the following.

\subsubsection{Milky Way Potential}

For this work, we use and modify the standard Milky Way potential contained within the \texttt{galpy} Python library \citep{Bovy_2015}, dubbed \texttt{MWPotential2014}. The models used in this potential were fitted with various data sources, the details of which are further elaborated in \citet{Bovy_2013} and \citet{Bovy_2015}. The potential consists of an NFW dark matter halo, a power-law bulge potential and a Miyamoto-Nagai disk potential \citep{miyamoto_nagai}. For the purposes of this work, we modify \texttt{MWPotential2014} as done in \citet{battaglia_edr3} by increasing the virial mass, $M_\mathrm{vir}$, of the dark matter halo to 1.6$\,\times 10^{12}\,M_\odot$, while also using the original virial mass of 0.8$\,\times 10^{12}\,M_\odot$ so that we have a `heavy' and `light' Milky Way potential, respectively, which are in the upper and lower range of virial masses estimated for the Milky Way in the literature \citep{Wang_2020}. We note, however, that this is not a fully realistic choice of the Milky Way potential for several reasons: it does not take the time evolution of the gravitational potential due to the Milky Way's assembly \citep{correa,Evans_2020} into account, nor the adiabatic contraction of the potential due to the accumulation of gas in the halo centre \citep{2020_cautun}, or asymmetries due to the rotating Galactic bar \citep{rotating_bar} or massive satellites \citep{Garavito_Camargo_2019}, although it is now well established that the Large Magellanic Cloud is likely to have strongly influenced the orbits of Milky Way's satellites \citep{patel_satellite_lmc_orbits,battaglia_edr3,correa_magnus}. It should also be noted that the \citet{Bovy_2015} potential has a very low dark matter fraction in the centre, which is incompatible with cosmological simulations in $\Lambda$CDM and may make it difficult to sustain a stable disk \citep{lovell_dm_fraction}. We also do not take dynamical friction into account in this work, as we found that it only results in minor deviations of the dwarf spheroidal orbits (the greatest effect is found in Fornax, where the orbit is delayed by $\sim$ 0.2 Gyr with the last apocentre offset by no more than 10 kpc, compared to the prediction of the Chandrasekhar approximation \citep{chandrasekhar} as implemented in \texttt{galpy}\footnote{We note that there is dynamical \textit{self-friction} present in our simulations, which also partially causes a decrease in dSph orbital apocentres.}). Although it is likely that our simulated satellite orbits do not follow the true ones, there is equally much uncertainty in the Milky Way mass \citep{Wang_2020}, the mass of the LMC \citep{lmc_mass, watkins_lmc}, whether the satellites had a past association with LMC \citep{santos_lmc} and how strongly the satellites are affected by dynamical friction due to their mass and the mass distribution in the Milky Way \citep{santistevan_mw_sats_a_mess}. These uncertainties would introduce an enormous set of parameters to be considered, and we thus focus on satellite orbits in a static Milky Way potential without including these additional unknowns. Taking the results of \citet{patel_satellite_lmc_orbits} and \citet{battaglia_edr3} at face value, this most likely means that our simulated satellites are affected by tides more strongly for `heavy' and `light' Milky Way potentials, as interaction with LMC favours a late infall for most of our satellites, although the cumulative effect of tidal forces to the present day may be somewhere between the `light' and `heavy' cases, particularly if the dwarfs had undergone tidal pre-processing as part of group infall \citep{genina_fornax,preprocess,callingham_group_infall}.

\subsubsection{Initial Conditions}

Following the approach of \citet{battagliaFornaxtides} and \citet{Iorio_2019}, who produced models of Fornax and Sculptor orbiting in the Milky Way potential in a setup similar to ours, we use an iteration procedure to converge on the initial properties of the dark matter halo and the stars, such that the stellar population has similar photometric and kinematic properties as observed in the dSphs today. We discuss this procedure further in Section \ref{sec:dsph_samples} and, in the following, we describe the initial guesses for the dark matter halo and stellar distribution parameters and how the dwarfs are set on an orbit around the Milky Way potential.

To generate initial conditions, we use \texttt{MAGI} \citep{Miki2017:1712.08760v1}, which utilises a distribution-function-based method and supports various density models. This method uses Eddington's formula \citep{edding} to produce the distribution function $f$, using the first and second derivative of the volume-density $\rho$ with respect to the system potential. \texttt{MAGI} also offers the option to provide a density profile cut-off radius $r_c$, which uses a complementary-error-based smoother. In this work, we use a Plummer profile for the stellar population in a dwarf spheroidal, as these profiles represent well stellar distributions in nearby dwarf galaxies \citep{moskowitz}, and an NFW dark matter halo.

For the initial masses of the dark matter haloes, we take the estimates from \citet{Read_2019}, who determine pre-infall virial masses through abundance matching with the mean star formation rate. We then use the mass-concentration-redshift relation in $\Lambda$CDM of \citet{Ludlow_2016} to obtain the concentration $c$ of the NFW halos\footnote{We take the concentration at z=0 rather than at the redshift of infall as the relation of \citet{Read_2019} is calibrated on inferred halo masses of isolated dwarfs at the present day. In any case, the concentration values will change as we use the iteration procedure to determine dark matter halo parameters that best fit the data in Section \ref{sec:dsph_samples}. }. The stellar masses for the 5 dwarfs we investigate are taken from \citet{de_Boer_2014}, \citet{de_Boer_2012a}, \citet{de_Boer_2012b}, \citet{Carrera_2002}, and \citet{Aparicio_2001}. As for the Plummer profile, we use the photometric data stored within the {\sc GravSphere} framework\footnote{ \url{https://github.com/justinread/gravsphere}} from \citet{carina_photo, koposov_2014} and \citealt{flewelling} (the data selection is described in section 4.2 of \citealt{Read2_2019}). Using this data, we fit normalised Plummer mass profiles to obtain the Plummer scale radius $a$, which we input in \texttt{MAGI}. 

 We use the virial and tidal radii as the exponential truncation radius for the NFW halo and as the cutoff for the Plummer profile, respectively\footnote{This choice is motivated by the fact that the outermost particles are quickly removed once the dwarf is placed in a Milky Way potential, and therefore this choice allows us to keep the resolution within the dwarf galaxy.}. For the tidal radius, we use the derivation in \citet{gal_dyn}:
\begin{equation}
    r_t = R_0\left(\frac{m(r_t)}{M(R_0)(3 - \frac{d\ln M}{d\ln R}\rvert _{R=R_0}} \right)^{1/3}\,,
    \label{eq:tidal_r}
\end{equation}
\\
where $M(R)$ is the enclosed mass of the host for a distance $R$ from its centre, $m(r)$ is the same but for the satellite and its centre, and $R_0$ is the orbital radius of the satellite from the centre of the host, which we take as the initial position of the dwarf with respect to the Milky Way in the initial conditions.

To set the dwarf spheroidals in orbit, we use \texttt{galpy} \citep{Bovy_2015} to obtain the 3-D positions and velocities of each dwarf in the `light' and `heavy' Milky Way potentials as a function of time. We then determine the infall time of each dwarf as the orbital apocenter nearest to the infall time estimates of \citet{Rocha_2012}\footnote{This was determined through the deduction of a correlation between orbital energy and the time at which dSphs last cross the Milky Way virial radius, found through cosmological simulations.}. Although this infall time estimate is derived from cosmological simulations, it nevertheless does not select Milky Way analogues with an LMC companion and thus our simulations are unlikely to represent the full tidal history of the dwarf sample. The position and velocity vectors at this point are used to initialise the orbit of each dwarf. We list the relevant orbital parameters in Table~\ref{tab:dsphs_gal_real}.

It should be noted that after initialisation with \texttt{MAGI}, the dSphs were left to simulate with GADGET-4 for 5 Gyr without an external potential for them to settle in equilibrium and minimize any numerical artefacts. Only thereafter were the dSphs `shifted' to the starting apocenter radius, had their velocities `boosted' into orbit, and left to evolve in heavy and light Milky Way potentials until their positions at the present day. We additionally note that since velocity anisotropy $\beta(r) = 0$ (isotropic) is disfavoured for cored stellar profiles living in cuspy dark matter haloes \citep{an_evans, almeida_cores_in_cusps}, the resultant stellar systems instead typically show radially varying velocity anisotropy, with a preference for mildly tangential anisotropy in the centre. We note that \citet{wolf_bullock}, using Jeans dynamical modeling with constant anisotropy, also find that Milky Way's dwarfs require slightly tangential anisotropies given their dark matter haloes have an NFW form.

\begin{table}
    \centering
    \setlength{\tabcolsep}{4pt} 
    \renewcommand{\arraystretch}{1.2} 
    \begin{tabular}{|l|c|c|c|c|c|}
        \hline
        \textbf{DSph} & \textbf{Fornax} & \textbf{Draco} & \textbf{Ursa Minor} & \textbf{Carina} & \textbf{Sculptor} \\ 
        \hline
        (h) $t_\mathrm{infall}$ [Gyr] & 8.77 & 9.05 & 9.96 & 9.07 & 7.74 \\
        (l) $t_\mathrm{infall}$ [Gyr] & 8.48 & 8.71 & 8.70 & 8.74 & 8.17 \\
        (est.) $t_\mathrm{infall}$ [Gyr] & 7 - 10 & 8 - 10 & 8 - 11 & 7 - 10 & 7 - 9 \\
        \hline
        (h) $r_\mathrm{apo}$ [kpc] & 154.9 & 90.4 & 87.6 & 107.8 & 94.4 \\
        (l) $r_\mathrm{apo}$ [kpc] & 206.1 & 125.0 & 109.4 & 221.8 & 128.0 \\
        (h) $r_\mathrm{peri}$ [kpc] & 80.8 & 30.9 & 30.3 & 88.3 & 39.8 \\
        (l) $r_\mathrm{peri}$ [kpc] & 132.1 & 44.5 & 45.6 & 106.9 & 59.2 \\
        \hline
        $r_{MW}$ [kpc] & 149 & 76 & 78 & 107 & 86 \\
        $v_{MW}$ [km/s] & -33 & -32 & -85 & 7 & -79 \\
        \hline
    \end{tabular}
    \caption{Infall times and orbital properties for the simulated dSphs. The infall times were estimated using \texttt{galpy} and compared to the estimated infall windows from \citet{Rocha_2012}. Orbital properties include apocenter and pericenter radii for heavy (h) and light (l) Milky Way potentials, as well as the Galactocentric distance and the Galactocentric velocity from \citet{McConnachie_2012}.}
    \label{tab:dsphs_gal_real}
\end{table}

\subsubsection{Evolution}

We utilize the $N$-body code GADGET-4 \citep{Springel_2021} to simulate the evolution of the dSphs. {\sc GADGET-4} also offers group finding methods, which use the classic friends-of-friends (FOF) approach to find groups of particles of approximately virial overdensity \citep{Springel_2001}, and the {\sc SUBFIND} algorithm to identify gravitationally bound substructures in these groups in configuration space \citep{subfind,Dolag_2009}. We use this to identify which particles get tidally stripped during the simulation and which remain bound.

To model the Milky Way potential, we modify {\sc GADGET-4} to read from a table of accelerations binned in cylindrical coordinates across the simulation volume. The acceleration at a specific coordinate is then calculated from linear interpolation between bins. This method was adapted from that of \citet{Tress_2020} within the {\sc AREPO} code \citep{Springel_2010}. We use \texttt{galpy} to construct the table of accelerations for the `light' and `heavy' Milky Way potential models. 

We simulate the dwarfs with $10^7$ particles in the dark matter halo and choose the number of stellar particles such that their masses are equal to the dark matter particle mass in order to avoid artificial spatial expansion of the stellar component of the dSph \citep{Ludlow_2019}. \citet{van_den_Bosch_2018} have shown that inadequate choice of the gravitational softening for a given number of particles can result in artificially enhanced tidal stripping and disruption of subhalos. We, therefore, calculate an ideal softening length for a subhalo with $N = 10^7$ particles based on \citet{van_den_Bosch_2018}, who find a relation for an NFW profile with corresponding concentration $c$ at which such instabilities can still be avoided, which we give here as:\\
\begin{equation}\label{eq:soften}
f_{\text {bound }}>\frac{1.79}{f(c)}\left(\frac{\varepsilon}{r_{\mathrm{s}, 0}}\right)\left(\frac{r_h}{r_{\mathrm{s}, 0}}\right),
\end{equation}
\\
where $f(c) =  \ln(1+c)-\frac{c}{1+c}$, $r_h$ is the half-mass radius of the halo, $r_{s,0}$ is the initial scale radius of the subhalo within the host galaxy, and $f_\mathrm{bound}$ is the fraction of particles that remain gravitationally bound after some time of evolution. To calculate this, we initially simulate the orbit of halos with a smaller number of particles ($10^6$) with a relatively low softening length of 50~pc and use the SUBFIND algorithm to find $f_\mathrm{bound}$ after 10 Gyr of evolution. From this, we use Equation \ref{eq:soften} to find the maximal softening length that fits the condition.

From all the dSphs of our study, we get a minimal softening length of $\epsilon = 7$~pc, which we use as a standard value for all simulated dwarfs. Along with this low softening length, Table \ref{tab:dsphs_res} highlights thus the high resolution used for this particular kind of study regarding the number of particles and their mass.

\begin{table}
    \centering
    \setlength{\tabcolsep}{4pt} 
    \renewcommand{\arraystretch}{1.2} 
    \begin{tabular}{|l|c|c|c|c|c|}
        \hline 
        \textbf{DSph} & \textbf{Fornax} & \textbf{Draco} & \textbf{Ursa Minor} & \textbf{Carina} & \textbf{Sculptor} \\ 
        \hline
        (h) m$_{\rm P}$ [$M_\odot$] & 1471.2 & 176.5 & 313.0 & 80.0 & 603.2 \\
        (l) m$_{\rm P}$ [$M_\odot$] & 1194.5 & 201.4 & 280.0 & 74.1 & 514.3 \\
        (h) N$_{\rm *}$ & 30452 & 1539 & 1082 & 4750 & 3431 \\
        (l) N$_{\rm *}$ & 34328 & 1632 & 1189 & 5152 & 4500 \\
        (h) N$_{\rm *}$ (today) & 29063 & 1458 & 908 & 4672 & 3298 \\
        (l) N$_{\rm *}$ (today) & 34091 & 1617 & 1159 & 5123 & 4463 \\
        \hline
    \end{tabular}
    \caption{Particle masses and counts used to simulate dSphs in the light (l) and heavy (h) Milky Way potential. The last two rows are the counts of stellar particles still bound to the dSphs at the present-day snapshot. Dark matter particle number was fixed at $10^7$ particles for all simulations and had the same mass as the stellar particles.}
    \label{tab:dsphs_res}
\end{table}

\subsubsection{Dwarf Spheroidal Galaxy Sample}\label{sec:dsph_samples}

\begin{figure}
    \centering
    \includegraphics[width=\columnwidth]{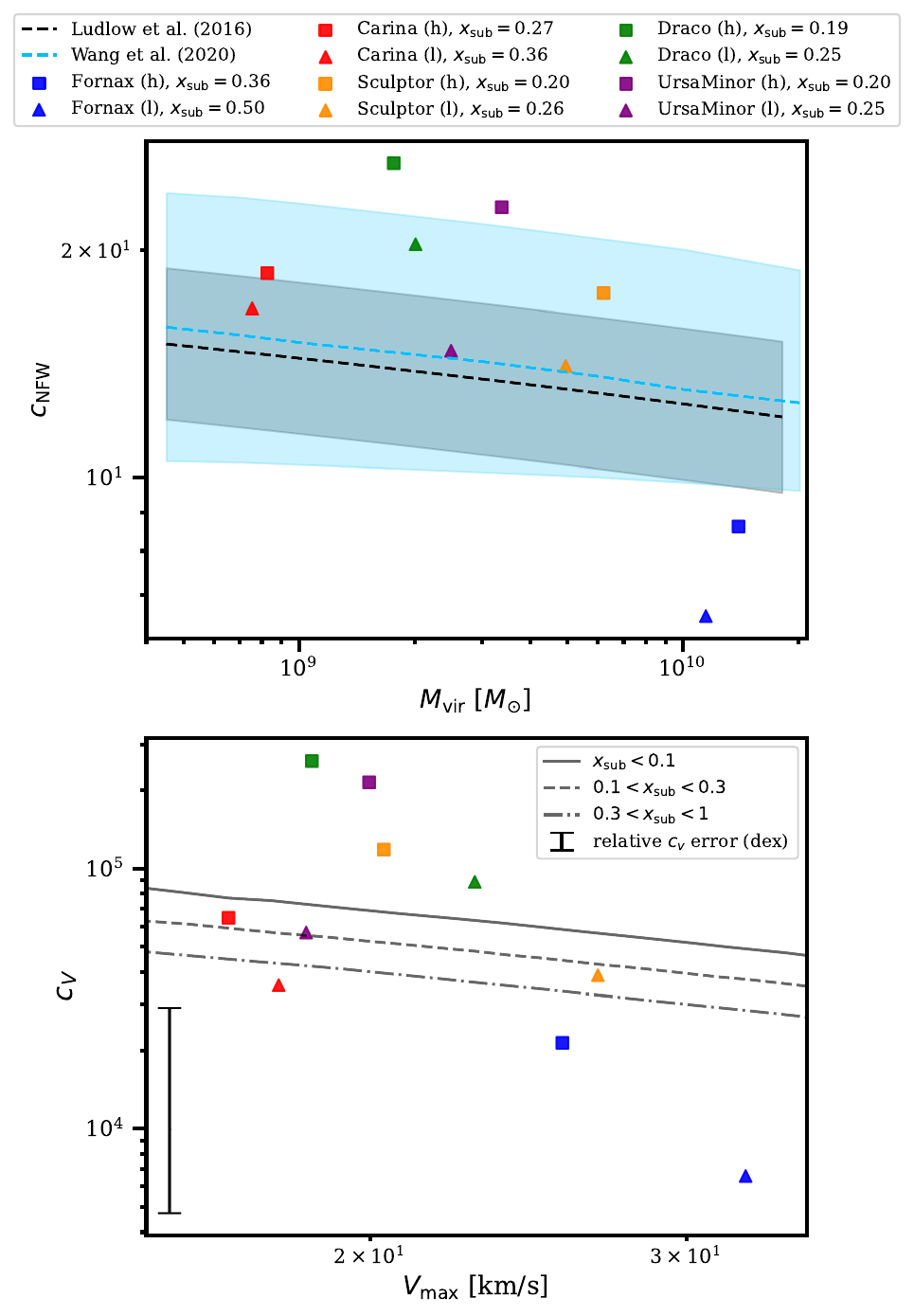}
    \caption{{\it Top:} initial masses and concentrations of dSphs in the light and heavy Milky Way potentials, after utilising an iterative procedure based on that from \citet{Iorio_2019}, which produces snapshots of dSphs at the present day that match the properties of dSphs from observations. The black dashed line and the shaded band are the mass-concentration relation at $z=0$ from \citet{Ludlow_2016} and the 0.1 dex errors \citep{dutton_maccio}. Dwarf galaxies in the `heavy' and `light' Milky Way potential are identified in the legend. The blue dashed line and shaded band show the mass-concentartion relation from the dark-matter-only simulations of \citet{wang_bose}. {\it Bottom:} the relation between the concentration parameter, $c_V$, and the maximum circular velocity, $V_{\rm max}$ measured for satellites from cosmological simulations in \citet{moline}. The grey curves show the average relation for subhaloes of a given $x_{\rm sub}$ range. Colour symbols show our sample of simulated dwarf spheroidals, for which the values of $x_{\rm sub}$ are shown in the legend. The black error bar shows the error in $c_V\approx0.28$~dex at the mass scale of Draco, calculated in \citet{massari_draco} and is representative of the typical error in the $c_V-V_{\rm max}$ relation in the mass range of our simulated dwarfs. }
    \label{fig:c_mass_comb}
\end{figure}

\begin{figure*}
    \centering
    \includegraphics[width=2\columnwidth]{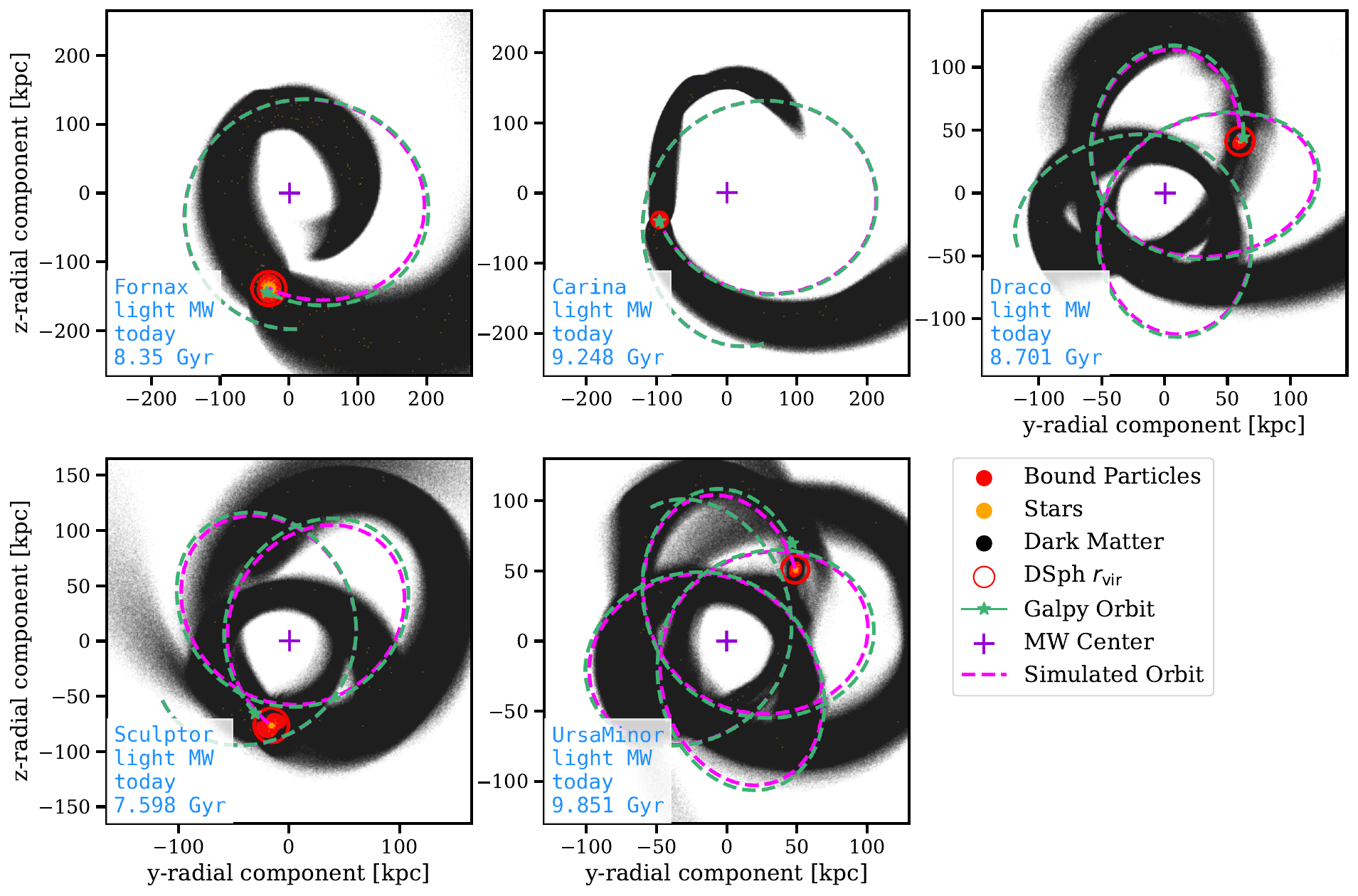}
    \caption{Snapshots in the `light' Milky Way potentials of the dSphs at present day. Black shows the distribution of dark matter and orange the stars. Bound particles, as identified with SUBFIND, are shown in red. Green is the expected (\texttt{galpy}) orbit and magenta is the orbit computed from the centre-of-mass of the bound particles of the dSphs. The Milky Way potential is not depicted here but is centred at the origin.}
    \label{fig:today_snaps}
\end{figure*}

\begin{figure*}
    \centering
    \includegraphics[width=2\columnwidth]{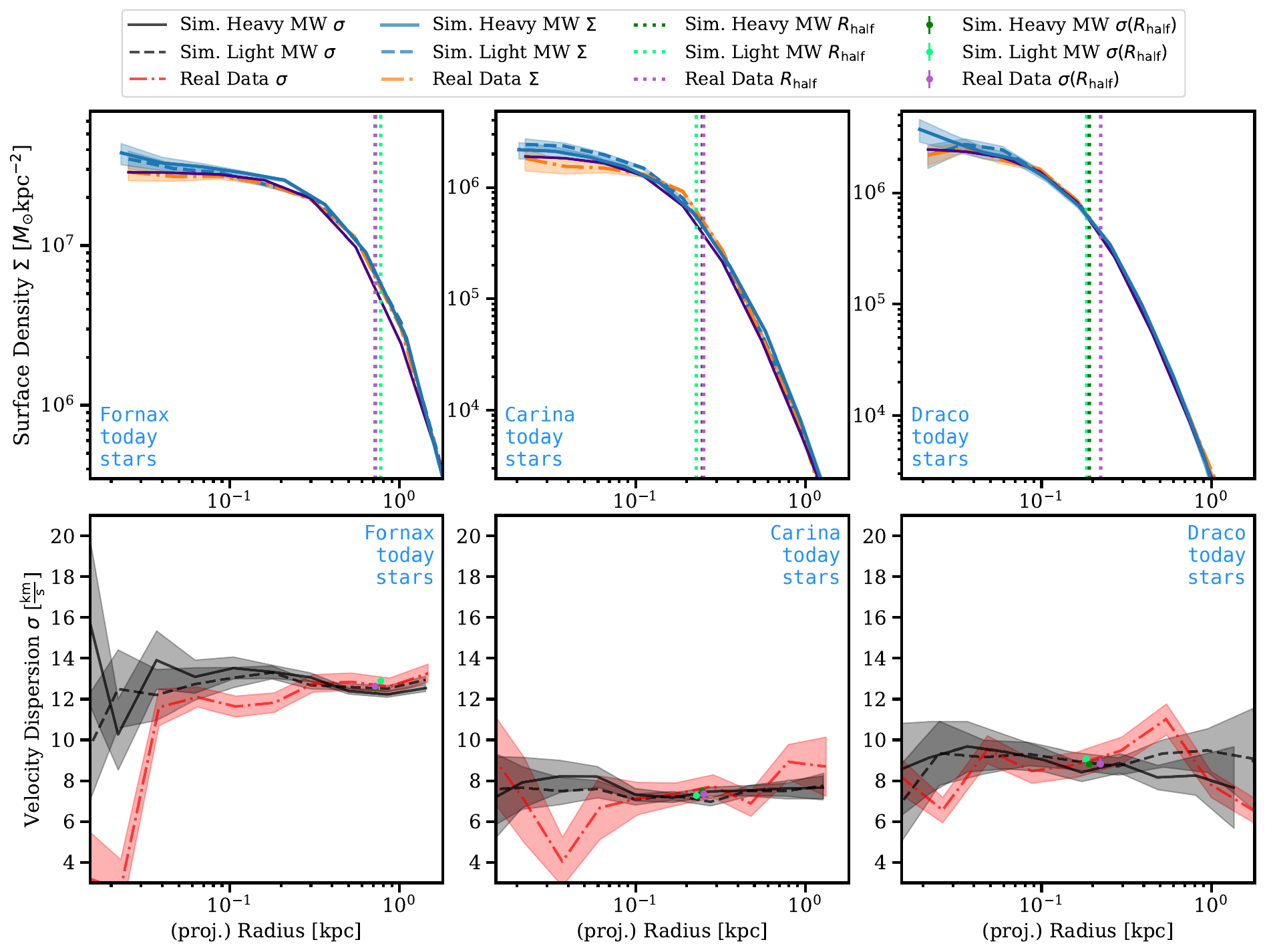}
    \caption{Comparison of the present-day surface density (top panels) and line-of-sight velocity dispersion profiles (bottom panels) for Fornax, Carina and Draco in the `heavy' (solid lines) and the `light' potential (dashed lines) compared to observational data (dot-dashed lines). The projected radius limit is determined by the extent of the observational data. The shaded bands correspond to Poisson errors in the surface density and Poisson errors combined with measurement errors for the line-of-sight velocity dispersion (2 kms$^{-1}$ measurement errors were assumed for the simulated data). Additionally shown here are the projected half-light radii obtained from fitting a Plummer profile and the corresponding line-of-sight velocity dispersion within the half-light radius from simulations (dotted dark and light green lines and error bars) and observations (purple dotted lines and error bars). The solid purple lines in the top panel show the best-fitting Plummer profiles to the observational data. The observational photometric and kinematic data in this figure and Fig.~\ref{fig:today_obs_comp2} were obtained from \citet{carina_photo, koposov_2014,flewelling,ursa_min_stellar,fornax_stellar,draco_stellar}.}
    \label{fig:today_obs_comp}
\end{figure*}

\begin{figure}
\centering
\includegraphics[width = \columnwidth]{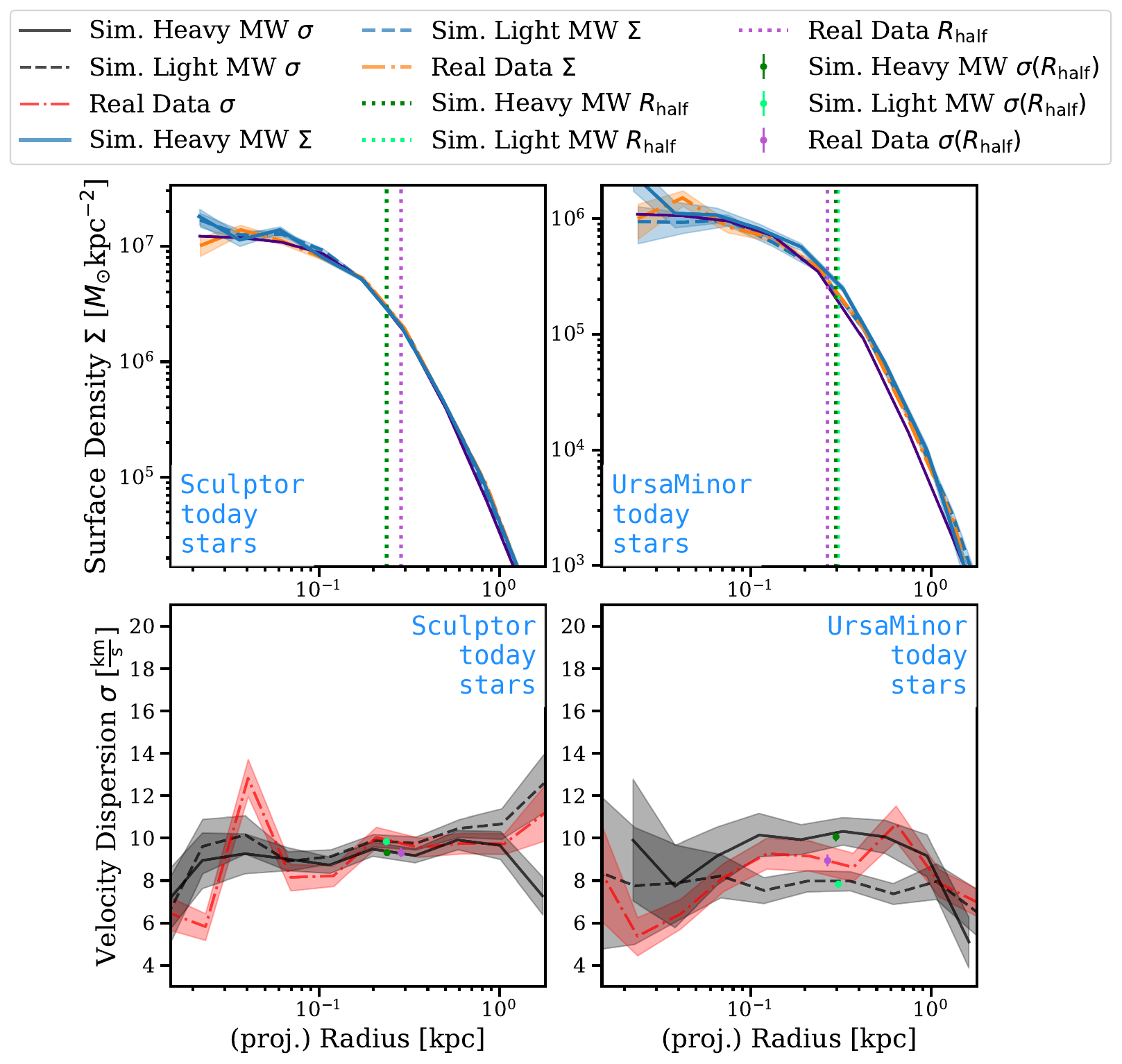}
\caption{Same as Fig.~\ref{fig:today_obs_comp}, but for Sculptor and Ursa Minor.}
\label{fig:today_obs_comp2}
\end{figure}

\begin{table}
    \centering
    \setlength{\tabcolsep}{4pt} 
    \renewcommand{\arraystretch}{1.2} 
    \begin{tabular}{|l|c|c|c|c|c|}
        \hline 
        \textbf{DSph} & \textbf{Fornax} & \textbf{Draco} & \textbf{Ursa Minor} & \textbf{Carina} & \textbf{Sculptor} \\ 
        \hline
        (h) $M_\mathrm{vir}$ [$10^9 M_\odot$] & 14.71 & 1.77 & 3.13 & 0.80 & 6.03 \\
        (l) $M_\mathrm{vir}$ [$10^9 M_\odot$] & 11.95 & 2.01 & 2.80 & 0.74 & 5.14 \\       
        (h) $r_\mathrm{cut}$ [kpc] & 39.8 & 19.7 & 23.8 & 15.1 & 29.6 \\
        (l) $r_\mathrm{cut}$ [kpc] & 37.2 & 20.5 & 22.9 & 14.7 & 28.1 \\
        (h) $r_\mathrm{s}$ [kpc] & 5.87 & 0.98 & 1.58 & 1.06 & 2.20 \\
        (l) $r_\mathrm{s}$ [kpc] & 7.22 & 1.31 & 1.73 & 1.14 & 2.55 \\
        
        \hline
        (h) $M_*$ [$10^5 M_\odot$] & 448.0 & 3.02 & 3.38 & 3.80 & 23.68 \\
        (l) $M_*$ [$10^5 M_\odot$] & 432.8 & 2.94 & 2.90 & 3.82 & 23.14 \\
        (h) $a$ [kpc] & 0.63 & 0.17 & 0.26 & 0.23 & 0.20 \\
        (l) $a$ [kpc] & 0.74 & 0.18 & 0.29 & 0.22 & 0.21 \\
        
        (h) $r_\mathrm{tidal}$ [kpc] & 25.4 & 8.3 & 9.8 & 9.5 & 12.9 \\
        (l) $r_\mathrm{tidal}$ [kpc] & 37.0 & 13.6 & 8.1 & 15.6 & 19.0 \\
        \hline
    \end{tabular}
    \caption{Structural parameters for the sample of dSphs. The table includes total virial and stellar mass in `heavy' (h) and `light' (l) Milky Way potentials. It also summarises key parameters for the NFW and Plummer profiles, including the NFW scale radius ($r_s$), the halo exponential cut-off radius ($r_{\mathrm{cut}}$), the Plummer scale radius ($a$) and the tidal radius at infall ($r_{\mathrm{tidal}}$).}
    \label{tab:dsphs_mass_params}
\end{table}

We selected a sample of dSphs based on their orbits, range of mass and availability of spectroscopic and photometric data. We wished to have variety in these aspects. Thus, we chose the following five dSphs: Fornax, Carina, Ursa Minor, Sculptor, and Draco.

In order to produce snapshots of dSphs that match their stellar masses and dynamics from observations, we use an iterative procedure from \citet{battagliaFornaxtides} and \citet{Iorio_2019} to adjust the initial virial masses, stellar masses, NFW scale radii, and Plummer scale radii. Table \ref{tab:dsphs_mass_params} summarises the information used to initialise each dSph with \texttt{MAGI}. The top panel of Figure \ref{fig:c_mass_comb} shows the resulting virial mass and concentration compared to the concentration-mass relation in $\Lambda$CDM of \citet{Ludlow_2016}. Interestingly, many of the simulated dwarfs, particularly in the `heavy' Milky Way potential, are outliers in the mass-concentration relation (although we note that the 0.1 dex scatter \citep{dutton_maccio} we show here may not be representative of scatter at small halo masses, as seen in \citealt{Ludlow_2016}). This may suggest that the infall times of \citet{Rocha_2012} occur too early for these dwarfs or that the mass of the Milky Way must be significantly lower than $1.6\times10^{10}M_{\odot}$ for dwarf galaxy properties to be consistent with $\Lambda$CDM. This could be an interesting avenue to explore for studies aiming to constrain the mass of the Milky Way. As we will show in Section \ref{sec:l v h}, the line-of-sight velocity dispersion profiles of dwarf galaxies may also be indicative of the Milky Way's mass.

The dashed cyan line and shaded band show the mass-concentration relation from the simulations of \citet{wang_bose}, who explore 20 orders of magnitude in halo mass, covering the range of expected halo masses of our simulated dwarfs. It can be seen that the scatter in this relation is substantially larger, easing the tension with the dwarfs in the `heavy' Milky Way potential, while `light' Fornax remains an outlier. We note, however, that the simulations of \citet{wang_bose} focus on the void regions of the $\Lambda$CDM universe and thus may not be applicable to the expected concentrations of dwarf galaxies in Local Group-like environments \citep{hellwing}.

The bottom panel of \ref{fig:c_mass_comb} shows the relation between the concentration parameter,

\begin{equation}
c_V = 2\left(\frac{V_{\rm max}}{H(z)R_{\rm max}} \right)^2  ,  
\end{equation}
and $V_{\rm max}$ from \citet{moline}, with $V_{\rm max}$ the maximum circular velocity occurring at the radius $R_{\rm max}$ and $H(z)$ the Hubble parameter at a given redshift $z$. The relation changes as a function of $x_{\rm sub}=R_{\rm sub}/R_{\rm vir}$, where $R_{\rm sub}$ is the distance of the subhalo from the host galaxy and $R_{\rm vir}$ is the virial radius of the host. The black error bar, shows the typical scatter in the relation in this range of $V_{\rm max}$ for a given $x_{\rm sub}$ from \citet{massari_draco}. Instead of individual $x_{\rm sub}$ values, we show the relation marginalised for ranges of x$_{\rm sub}$ for clarity. In agreement with the mass-concentration relation in the upper panel, Draco and Ursa Minor in the `heavy' potential, as well as Fornax in the `light' potential are outliers in the relation, independent of $x_{\rm sub}$.

The orbital trajectories of our sample of dwarf spheroidals in the `light' Milky Way potential up to the present day are shown in Figure \ref{fig:today_snaps}. The dashed purple line highlights the orbit of the dwarfs in the simulation, while the dashed green line shows the orbit predicted by {\sc galpy}. It can be seen that the two orbits agree rather well; however, due to the dynamical self-friction of the dwarfs by their own debris \citep{Miller_2020}, there are minor offsets for the more massive dwarfs that are particularly noticeable at the orbital apocentre. It can also be seen that our simulated dwarfs can lag behind the orbits predicted by {\sc galpy}. In our dynamical modelling of the dwarfs at the present day, we always pick the simulated dwarf at the correct orbital phase that corresponds to the {\sc galpy} prediction, quantified by the correct position of the dwarf with respect to the Milky Way's disk and the direction of its velocity vector.

In Figure~\ref{fig:today_obs_comp} and~\ref{fig:today_obs_comp2}, we show the resulting line-of-sight velocity dispersion and surface density profiles for each dwarf from our simulations and real data from \citet{fornax_stellar, draco_stellar, ursa_min_stellar, carina_photo, koposov_2014, flewelling}. We note that throughout this work, we place the ``observer'' at infinity from the direction of the Milky Way's centre rather than at $\sim$8~kpc in the Galactic disk. We have made this choice as we explore various orbital phases of the dSphs in this work, with the dynamics and tidal stripping determined primarily by the motion of the dwarf with respect to the Galactic centre. We thus prefer to remove the ambiguity of Solar motion around the Galaxy and the uncertainties associated with the spatial extent of dwarf galaxies on the sky. Since we explore initially spherical NFW halos, which at the present day lie at least 76~kpc from Milky Way's centre (corresponding to at most 6~degree offset from the Solar position), this choice does not significantly affect the constrained line-of-sight properties of our simulated dwarf galaxies or the results of this work, which we have also explicitly verified. 

In the surface density, our simulated dwarfs are in good agreement with the observational data, particularly in the `light' potential, with a notable exception of Carina, where the observational data show a nearly flat surface density in the central 200~pc, which is not well described by a Plummer profile (purple line). The Fornax `heavy' model also overestimates the surface density within the central 300~pc. Fornax is known to not be well-fit by a single Plummer profile, since the outer profile falls off more steeply than allowed by the model \citep{moskowitz}. In Ursa Minor, a single Plummer model also struggles to describe the profile shape, although our `light' model is in good agreement with the observed surface density. 

In the line-of-sight velocity dispersion, the agreement is again better in the `light' than the `heavy' potential and varies as a function of radius. For Fornax, the velocity dispersion in both cases is overestimated in the central 300~pc, while the value within the half-light radius is in good agreement between the model and observations. 

 From the agreement between the half-light radius and the velocity dispersion calculated within, it is clear that the iterative procedure worked well for all the dSphs except Ursa Minor, which could not converge to the observed velocity dispersion despite numerous iterations in the light and heavy Milky Way potentials. This dSph is the most tidally affected in our sample and, as pointed out by \citet{battagliaFornaxtides}, in the regime where the structure of the dwarf changes significantly due to tidal stripping and tidal shocks, the iterative algorithm may not be able to converge. This is because the resultant line-of-sight velocity dispersion and the half-light radius may not scale linearly with the changes in the dark matter halo and stellar parameters in each new parameter iteration. Convergence in this case is also likely affected by the stochasticity in each newly generated set of initial conditions, as the number of stellar particles is particularly low ($\sim 10^3$).  

 \citet{battagliaFornaxtides} modelled Fornax using a cored dark matter halo and an approximate deprojection of a \citet{sersic} surface density profile for the stars. Their model of surface density profile provided a better fit in the central regions of Fornax, which flattens towards the centre more quickly than the Plummer model; however, their recovered line-of-sight velocity dispersion profiles are comparable to ours, with difficulty in reproducing the lower velocity dispersion in the central 0.5~kpc.

\citet{Iorio_2019} modelled Sculptor with a Plummer law, embedded in a cored NFW-like halo. Their models seem to slightly overestimate the surface density in the inner regions (< 100~pc) and underestimate it in the outer regions, while our models slightly overestimate the surface density around 50~pc. The line-of-sight velocity dispersion profiles in their models provide a similar quality of fit to the data as we find here.

\subsection{Dark Matter Distribution Inference}
\label{gravspheresection}

A popular method to infer the dark matter distribution in dwarf galaxies has been the Jeans analysis. In the most common form, this makes use of the spherical Jeans equation, derived from the collisionless Boltzmann equation under the assumption of spherical symmetry and the distribution function that is constant in time \citep{binney82}:
\\
\begin{equation}\label{eq-spher-jeans-anisotropy-mass}
    M(<r) = -\frac{r\,\sigma_r^2}{G}\,\left(\frac{\mathrm{d} \ln (\nu\,\sigma_r^2)}{\mathrm{d} \ln r} + 2\,\beta\right)\,.
\end{equation}
\\
The equation relates the enclosed mass $M(<r)$ of the entire system to the radius $r$, radial velocity dispersion $\sigma_r$, tracer density $\nu$, and the anisotropy parameter $\beta$, as well as Newton's gravitational constant $G$. The anisotropy parameter $\beta$ measures the ratio of the velocity dispersion in different directions, specifically that of the radial velocities $v_r$ and the velocities tangential to this ($v_\phi$, $v_\theta$), with $\beta \equiv 1 - \frac{\sigma_\theta^2 + \sigma_\phi^2}{2\,\sigma_r^2}$.
\\
Thus, Equation~\ref{eq-spher-jeans-anisotropy-mass} forms the basis for dark matter distribution inference from stellar matter, using vital assumptions of spherical symmetry and dynamical equilibrium.
\\

For classical dSphs orbiting the Milky Way, large sets of kinematic and photometric data have already been obtained, though mostly limited to line-of-sight velocities and surface densities. It is possible to recover the 3-dimensional density $\nu(r)$ from the measured projected surface density $\Sigma(R)$. This can be done through an Abel transform:
\\
\begin{equation}\label{eq-3ddens-2ddens}
    \nu(r) = -\frac{1}{\pi}\,\int_r^\infty\mathrm{d}R\,\frac{\mathrm{d} \Sigma(R)}{\mathrm{d} R}\,\frac{1}{\sqrt{R^2-r^2}}\,.
\end{equation}
\\
Furthermore, a relation between the projected line-of-sight velocity dispersion and surface density and their deprojected counterparts was derived from \citet{binney82}, which takes the form:
\\
\begin{equation}\label{eq-surfdens-siglos-jeans}
    \Sigma(R)\,\sigma_\mathrm{los}^2(R) = 2\,\int_R^\infty\,\mathrm{d}r\,\left(1-\beta(r)\,\frac{R^2}{r^2}\right)\,\frac{\nu(r)\,\sigma_r^2\,r}{\sqrt{r^2-R^2}}\,.
\end{equation}
\\
With Equations \ref{eq-spher-jeans-anisotropy-mass}, \ref{eq-3ddens-2ddens}, \ref{eq-surfdens-siglos-jeans} one thus has a system of equations to retrieve a total mass profile from a kinematic tracer population, through which one can assume a parameterization of a mass profile $M(<r)$ and anisotropy profile $\beta(r)$, and perform a best-fit for their parameters. In effect, the predicted $\sigma_\mathrm{los}(R)$ and $\Sigma(R)$ (both of which are dependent on the projected radius $R$) are compared with the kinematic and photometric data during the fit. However, since typically only line-of-sight information of the tracers is known, $\beta$ becomes poorly constrained. Thus, the Jeans dynamical modelling suffers from the mass-anisotropy degeneracy. 

Several methods have been explored to break this degeneracy. However, we use a method that exploits the fourth-order projected virial theorem (\citealt{merrifield, richardsonfairbairncusp,richardsonfairbairn2013, richardson,Read_2017}; see also \citet{hayashi_4order} for fourth-order Jeans equations), which utilizes two new equations for the so-called virial shape parameters (VSP):

\begin{equation}\label{eq:vsp1}
\mathrm{VSP1}=\frac{2}{5} \int_0^{\infty} G M v(5-2 \beta) \sigma_r^2 r d r=\int_0^{\infty} \Sigma\left\langle\sigma_{los}^4\right\rangle R d R
\end{equation}
and
\begin{equation}\label{eq:vsp2}
\text { VSP2 }=\frac{4}{35} \int_0^{\infty} G M v(7-6 \beta) \sigma_r^2 r^3 d r=\int_0^{\infty} \Sigma\left\langle\sigma_{los}^4\right\rangle R^3 d R.
\end{equation}
These equations use the fourth moment of the line-of-sight velocities $\left\langle\sigma_{los}^4\right\rangle$. They thus allow one to put two additional constraints on $\beta$, but may not fully break the mass-anisotropy degeneracy.

\subsubsection{pyGravSphere}

This work performs such a fit for the mass profile of dSphs with line-of-sight data using a Python framework called {\sc pyGravSphere}\footnote{\url{https://github.com/AnnaGenina/pyGravSphere}} \citep{Genina_2020}, which was adapted from the original {\sc GravSphere} code \citep{Read_2017}. It is based on the ensemble sampler {\sc EMCEE} \citep{Foreman_Mackey_2013}. 

We use the broken power-law option for the density profile, as done in \citet{Read_2017}, \citet{read_draco} and \citet{Read2_2019}. This uses five spatial bins defined as logarithmically spaced fractions of the projected half-light radius $R_e$, with the values $r_j = [0.25,0.5,1,2,4]R_e$, where $R_e$ is fixed to the scale radius of a single Plummer profile fit. Within each bin, the profile follows a power law defined by slopes with parameters $\gamma_j$. The complete profile is thus defined as:
\begin{equation}\label{eq:BPL}
\rho_{\mathrm{dm}}(r)= \begin{cases}\rho_0\left(\frac{r}{r_0}\right)^{-\gamma_0}, & r<r_0 \\ \rho_0\left(\frac{r}{r_{j+1}}\right)^{-\gamma_{j+1}} \prod_{n=0}^{n<j+1}\left(\frac{r_{n+1}}{r_n}\right)^{-\gamma_{n+1}}, & r_j<r<r_{j+1}\end{cases}
\end{equation}
with $\rho_0$ being the density at $r_0$. Beyond the last bin, the power law is extrapolated. This radial extent is sufficient for calculating J-factors later on, and it should also be noted that it covers typical ranges of kinematic tracers. We also fit the stellar surface density with a combination of 3 Plummer profiles \citep{plummer} (see Equation~\ref{eq:surf_dens_plum}) and assume a radially varying $\beta$, described by a function from \citet{Baes_2007}:
\begin{equation}
\beta(r)=\beta_0+\left(\beta_{\infty}-\beta_0\right) \frac{1}{1+\left(\frac{r_a}{r}\right)^\eta},
\end{equation}
where $\beta_0$ is the central value of the anisotropy profile, $\beta_{\infty}$ the value at infinity, $r_a$ the transition radius, and $\eta$ its steepness. In the {\sc pyGravSphere} priors, as shown in Table \ref{tab:priors}, we reparametrise the anisotropy as in \citet{Read_2017}:

\begin{equation}
     \Tilde{\beta} = \frac{\beta}{2-\beta}\,
 \end{equation}

allowing one to go from the range $-\infty < \beta < 1$ to a range of $-1 < \Tilde{\beta} < 1$, which is more easily sampled. It should be noted that {\sc pyGravSphere} takes into account the contribution of stellar mass to the gravitational potential. 

As described in the previous section, this framework effectively attempts to fit the line-of-sight velocity dispersion $\sigma_\mathrm{los}(R)$ and projected surface density $\Sigma(R)$, and also the two VSPs as from equations \ref{eq:vsp1} and \ref{eq:vsp2} by minimizing a log-likelihood function described as the sum of the $\chi^2$ of these four components:
\begin{equation}\label{eq:loglike}
    \ln{\mathcal{L}} = -\frac{1}{2}\left(\chi^2_{\sigma_\mathrm{los}} + \chi^2_\Sigma +\chi^2_{\mathrm{VSP1}}+\chi^2_{\mathrm{VSP2}}\right).
\end{equation}

In summary, we employ the same set of models and priors described in \citet{Read2_2019} for the dynamical modelling of our simulated sample of dwarf spheroidals. We list the priors in Table~\ref{tab:priors}. 

\begin{table}
\centering
\begin{tabular}{|c|c|c|}
\hline
\textbf{Parameter} & \textbf{Min} & \textbf{Max} \\ \hline
$\log_{10}\rho_0$ & 5 & 10 \\ \hline
$\gamma_0$ & 0 & 3 \\ \hline
$\gamma_1$ & 0 & 3 \\ \hline
$\gamma_2$ & 0 & 3 \\ \hline
$\gamma_3$ & 0 & 3 \\ \hline
$\gamma_4$ & 0 & 3 \\ \hline
$\tilde{\beta}_0$ & -1 & 1 \\ \hline
$\tilde{\beta}_\infty$ & -1 & 1 \\ \hline
$r_a$ & 0.5$R_e$ & 2$R_e$ \\ \hline
$\eta$ & 1 & 3 \\ \hline
$a_{1,2,3}$ & 0.5$a_{1,2,3}$ & 1.5$a_{1,2,3}$ \\ \hline
$M_{1,2,3}$ & 0.5$M_{1,2,3}$ & 1.5$M_{1,2,3}$ \\ \hline
$M_*$ & 0.75$M_*$ & 1.25$M_*$ \\ \hline
\end{tabular}
\caption{Priors for the parameters used by {\sc pyGravSphere}. The MCMC chains are initialised with a uniform distribution in the ranges shown. An additional parameter $\gamma_{\rm smooth}$ controls the maximum transition between consecutive density slopes and is set to 1.}
\label{tab:priors}
\end{table}

Finally, we point out that unlike the works of \citet{battagliaFornaxtides} and \citet{Iorio_2019}, we chose to carry out Jeans analysis on the kinematic sample of bound stars as determined by {\sc SUBFIND}, as opposed to a more observationally motivated kinematic sample selection. This is because we aim to study the departure of the orbiting dwarfs from dynamical equilibrium as a result of tides, as opposed to the effects of contamination of kinematic samples by tidally stripped stars, which has been previously explored in the literature \citep{klimentowski, klimentowski2, lokas_tides}; however, as we make our dwarf models publicly available it would be straightforward to incorporate these effects in future investigations.

\subsection{Wolf Mass Estimator}
\label{wolfsection}
For many dwarf satellites of the Milky Way, particularly the ultra-faints, large spectroscopic samples are often unavailable. In these cases, accurate estimates of mass enclosed within the half-light radius can be obtained with simple estimators \citep{walker_estimator}. One such method is the Wolf mass estimator \citep{Wolf_2010}, which is largely insensitive to the anisotropy parameter $\beta$. The estimator requires only the line-of-sight velocity dispersion and the (projected) half-mass radius, which is often available even with small stellar data samples. In this work, we therefore also want to test the accuracy of this estimator for a sample of satellites at different stages of their orbit, which could then inform the application of the \citet{Wolf_2010} mass estimator to other dwarfs.

For the `Wolf analysis', we use two forms of the \citet{Wolf_2010} mass estimator, the first of which we call the `3-D Wolf' estimator:
\begin{equation}\label{eq:wolf_3d}
    M(<r_3) \approx \frac{3\,\sigma^2_\mathrm{tot}(r_3)\,r_3\,}{G}\,,
\end{equation}
where $\sigma_\mathrm{tot}^2 = \sigma_r^2 + \sigma_\theta^2 + \sigma_\phi^2 = (3-2\,\beta)\,\sigma_r^2$, incorporating the full 3-dimensional kinematic information from the stars, and $r_3$ being a specific radius which fulfils:
\begin{equation}\label{eq-wolf-req-3}
\frac{\mathrm{d} \ln \nu}{\mathrm{d} \ln r}\Bigg|_{r_\mathrm{3}} = -3\,,
\end{equation}
and at which there is an insensitivity to the anisotropy parameter $\beta$.

In the absence of full 3-dimensional information, one can still make further approximations to Equation~\ref{eq:wolf_3d} to obtain the `2-D Wolf' estimator:
\begin{equation}\label{eq:wolf_2d}
    M(<r_{1/2}) \approx \frac{4\,\langle \sigma^2_\mathrm{los} \rangle\,R_e}{G}\,,
\end{equation}

with the luminosity-weighted average line-of-sight dispersion \citep{kowalczyk_2014}:
\begin{equation}\label{eq:lum_w}
\left\langle\sigma_\mathrm{los}^2\right\rangle=\frac{\int_0^{\infty} \sigma_\mathrm{los}^2(R) \Sigma(R) R d R}{\int_0^{\infty} \Sigma(R) R d R}\,.
\end{equation}

We develop an analysis pipeline to obtain Wolf mass estimates (which we dub the `Wolf analysis') for every snapshot of the dSph orbit (separated by 50~Myr). In this analysis, we obtain two different Wolf estimates: The 3-dimensional estimation of the Wolf mass using Equation \ref{eq:wolf_3d}, and the 2-dimensional projected Wolf mass using Equation \ref{eq:wolf_2d}. We estimate this also for two tracer components: the stars and the dark matter, with the latter informing us of how the estimator would perform if the kinematic tracers extended out to the tidal radius of the dark matter halo. We then compare with the `true' masses obtained directly from the simulation within $r_3$ and $r_{1/2}$ of the stars and the dark matter.

In the 3-D case, we construct the tracer density $\nu(r)$ and $\sigma_{\rm tot}(r)$ profiles by radially binning the stars and dark matter in bins of equal particle numbers (100 particles). The profiles are further smoothed with a cubic spline. The radius $r_3$ is obtained from deriving the log-slope profile $\frac{d\ln\nu}{d\ln r}$ and $\sigma_{\rm tot}(r3)$ is then straightforwardly found.

In the 2-D case, we fit a Zhao \citep{Zhao_1997} profile with a tidal modification to the projected dark matter mass distribution, similar to \citet{Read_2019}:
\begin{equation}
    \Sigma_{\mathrm{DM}}(R)=
    \begin{cases}
        \frac{\Sigma_s}{\left( \frac{R}{R_s} \right)^{\gamma} \left[ 1 + \left(\frac{R}{R_s}\right)^{\alpha} \right]^{\frac{\beta - \gamma}{\alpha}}}  & R \leq r_t\\
        \Sigma_{\rm DM}(r_t)\,\left(\frac{R}{r_t}\right)^{-\delta} & R > r_t\,,
    \end{cases}
\end{equation}
where $R_s$ and $\Sigma_s$ are the scale radius and density, $\gamma$, $\beta$ and $\alpha$ are the inner outer and transition slopes, respectively, and  $\delta$ is the slope of the tidal modification at radius $r_t$, which is typically the tidal radius, but in our case, it is left as a free parameter. Similarly, for the stars, we fit a Plummer surface mass profile with:
\begin{equation}
    \Sigma(R) = M_0\frac{a^2}{\pi(a^2 + R^2)^2}\,,
    \label{eq:surf_dens_plum}
\end{equation}
with $a$ being the scale radius and $M_0$ the total mass. We additionally use the same tidal modification for this profile as done for the Zhao profile.

From these fitted profiles, we obtain the projected half-mass radius\footnote{For the Plummer profile, we use the best-fit scale radius $a$, as the tidal modification is usually not utilized in the fit and is not converged, sometimes resulting in diverging mass profiles.} and the profiles are further used to compute the luminosity-weighted line-of-sight velocity dispersion via Equation~\ref{eq:lum_w}.  

Both the `Wolf analysis' and dynamical modelling with {\sc pyGravSphere} are performed exclusively on the \emph{bound} particles, as identified with {\sc SUBFIND}. Any particles that are tidally stripped (identified as unbound) in any given snapshot are ignored in the analysis of future snapshots. Aside from these constraints, we use all bound particles in our analysis. We note that this at least doubles the spectroscopic sample used in \citet{Read2_2019} for poorer resolved systems like Ursa Minor, but is comparable to samples expected in the near future from e.g. Prime Focus Spectrograph on Subaru \citep{pfs_dwarfs}. In modelling with {\sc pyGravSphere}, where the velocity dispersion profiles can be particularly finely binned, with $N/\sqrt{N}$ binning, we pay additional attention to the line-of-sight velocity dispersion profiles, ensuring that no data points lie outside the tidal radius of the galaxy and are not significant outliers from the main body of the galaxy. 

\subsection{J-Factor Calculation}
\label{sec:jfac_theory}
The J-factor is an essential parameter for dark matter annihilation searches. This value is proportional to the squared dark matter density, $\rho^2$, and is typically calculated like so for extended objects in the sky:

\begin{equation}\label{eq:j_factor_l}
J(\Delta\Omega) = \int_{\Delta\Omega}\int_\text{LOS}\rho^2(l,\Omega')\,\text{d}l\,\text{d}\Omega',    
\end{equation}
\\
with $l$ the distance along the line-of-sight, which runs from 0 to $\infty$, and $\Delta\Omega$ is the solid angle over which we integrate. In our study, we fix this to 0.5 degrees as done in most J-factor studies and corresponds for most dwarfs to a radius likely to contain most of the annihilation signal \citep{Calore_2018}. 

In this work, we calculate J-factors in two ways. The first is through evaluating Equation~\ref{eq:j_factor_l} with the density profiles\footnote{To re-parameterize the density distribution, $\rho(r)$, in terms of $l$ and $\Omega$ (or in the case of spherical symmetry only $\theta$), we use the method of \citet{Geringer_Sameth_2015}.} recovered by {\sc pyGravSphere} (i.e. observationally-derived J-factors). The second method computes the ``true" J-factor, where the density of the simulated dSph is estimated by calculating the local density of each bound dark matter particle. For each particle $i$ we find $N=32$ nearest neighbours 
 and calculate the density $\rho_i$ inside of a sphere containing all neighbours, centred on the particle $i$. For the density estimation, we select only the particles within the 0.5$^{\circ}$ cone starting from the observer and extending to a distance D to the dwarf. To optimize the search for the $N$ nearest neighbours, we use a KD-tree method \citep{10.1145/361002.361007}. The densities are then summed inside the solid angle, using the following equation to approximately obtain the final ``true" J-factor \citep{kuhlen}:

\begin{equation}\label{eq:J_approx}
    J(\Delta \Omega) \approx \sum_i \frac{\rho_i\,m_{DM}}{D^2}\,,
\end{equation}

where $m_{DM}$ is the mass of the dark matter particle in the simulation. 

When comparing the two methods, we keep the distance to the dSph constant (i.e. not necessarily the distance as measured in the simulation due to minor orbital deviations), taking values from the tabulated data in \citet{McConnachie_2012} who takes data from \citet{Carrera_2002}, \citet{Pietrzy_ski_2009}, \citet{fornax_stellar}, and \citet{bonanos_2003}.

\section{Results}
\label{results_section}

\subsection{Wolf Mass Estimator}\label{sec:wolf}

\begin{figure*}
    \centering
    \includegraphics[width=\columnwidth]{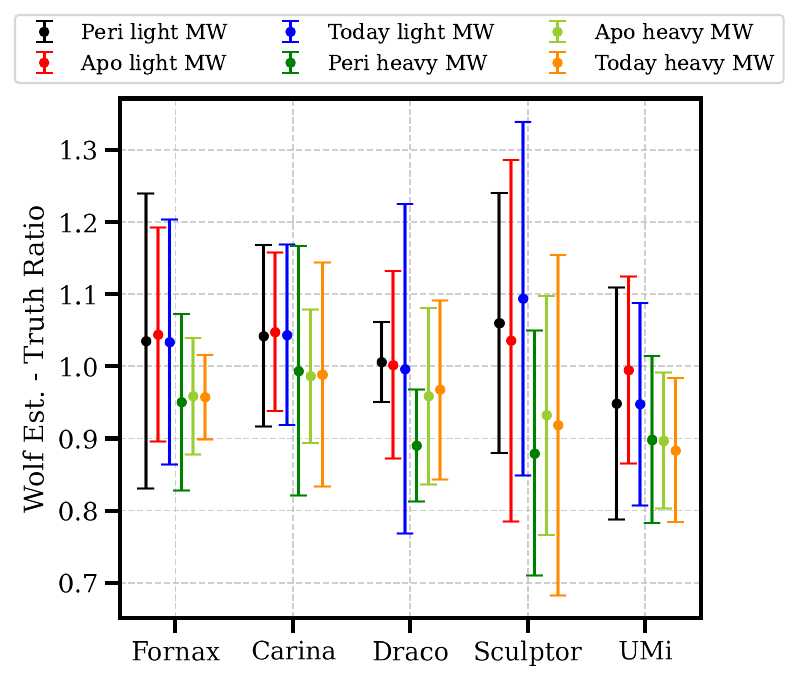}
    \includegraphics[width=\columnwidth]{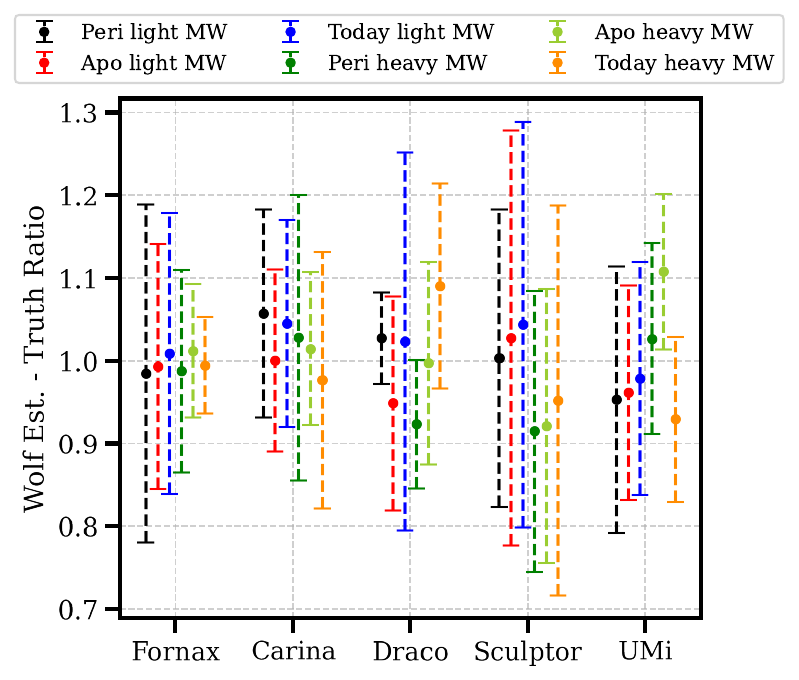}
    \caption{\textit{Left:} Ratios of the masses within the projected half-mass radii of the stellar populations computed with the `2-D Wolf' estimator  with the true masses within the 3-D half-mass radii. The ratios are shown for the present-day snapshots as well as nearest apocentre and pericentre snapshots in both the `light' and `heavy'Milky Way potentials. We average over all $\pm150$~Myr snapshots at each point. The errors were estimated from the Poisson error on the velocity dispersion within the half-light radius and the error on the Plummer profile $a$ fit, as well as the Poisson error on the mass enclosed within the 3-D half-mass radius. \textit{Right:} The same as the left, but the 2-D Wolf masses are instead computed using the half-mass radius directly obtained from the cumulative mass distribution of the bound particles (using projected radii) and using the mean line-of-sight velocity dispersion within the projected half-mass radius.}
    \label{fig:wolf_ratios}
\end{figure*}

On the left of Figure \ref{fig:wolf_ratios}, we summarise the results of the application of the 2-D Wolf mass estimator on our sample of dwarfs in the 2 Milky Way potentials, obtained through fitting surface brightness profiles to our dwarfs. We display the results for the present day as well as the nearest peri- and apocentres. It is clear that in the `light' potential, the mass within the half-mass radius is typically overestimated, while in the `heavy' potential the mass is typically underestimated by $\sim$5-10~per~cent. Exceptions to this include Ursa Minor, where both in the `heavy' and the `light' potential, the mass is underestimated. For mass estimates at the present day, the true value lies within the 68~per~cent confidence levels in both potentials for all dwarfs but Ursa Minor in the heavy potential. 

We compare these results with those using an alternative procedure of obtaining the projected half-mass radii directly from the cumulative distribution of bound mass. We also use the mean line-of-sight velocity dispersion within this half-mass radius instead of the numerically interpolated (and integrated) luminosity-weighted average line-of-sight dispersion as in Equation \ref{eq:wolf_2d}. The results are shown on the right of Figure~\ref{fig:wolf_ratios}. The method improves the performance of the estimator in some cases (Fornax), but also appears to disagree with the luminosity-averaged estimator, particularly at apocentres, with higher mass estimates in the heavy potential and lower mass estimates in the light potential.

While the results of the estimator are consistent within their errors for all orbital phases of the dwarfs, it can be seen that the overall accuracy of the estimator can change with orbital phase. To understand this result, we examine the extreme case of Ursa Minor in a `heavy' Milky Way potential over the whole course of its orbital history, applying the Wolf estimator to both dark matter and stars used as tracers.
 
\begin{figure}
    \centering
    \includegraphics[width=0.95\columnwidth]{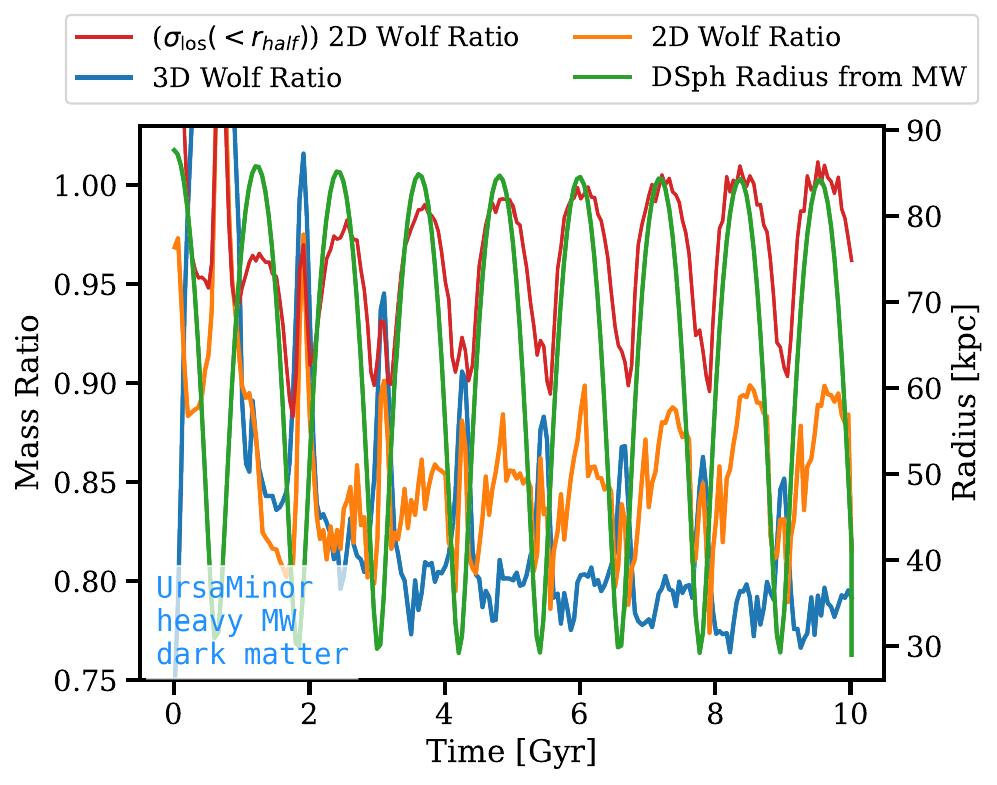}\\
    \includegraphics[width=0.95\columnwidth]{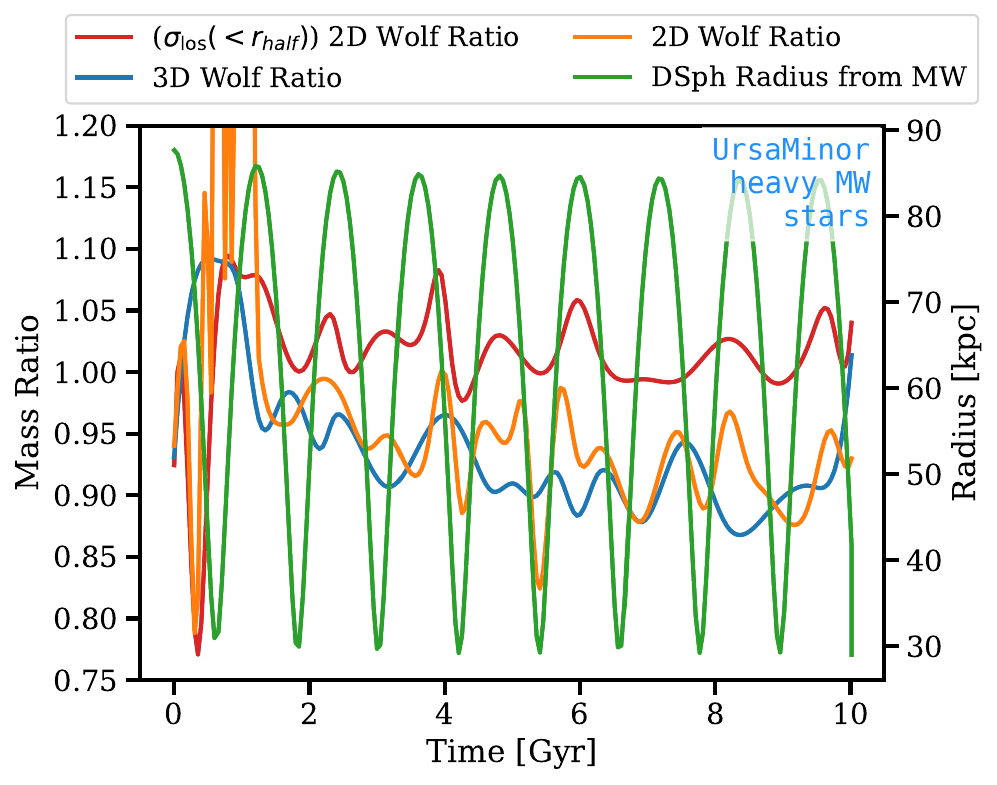}\\
     \includegraphics[width=0.95\columnwidth]{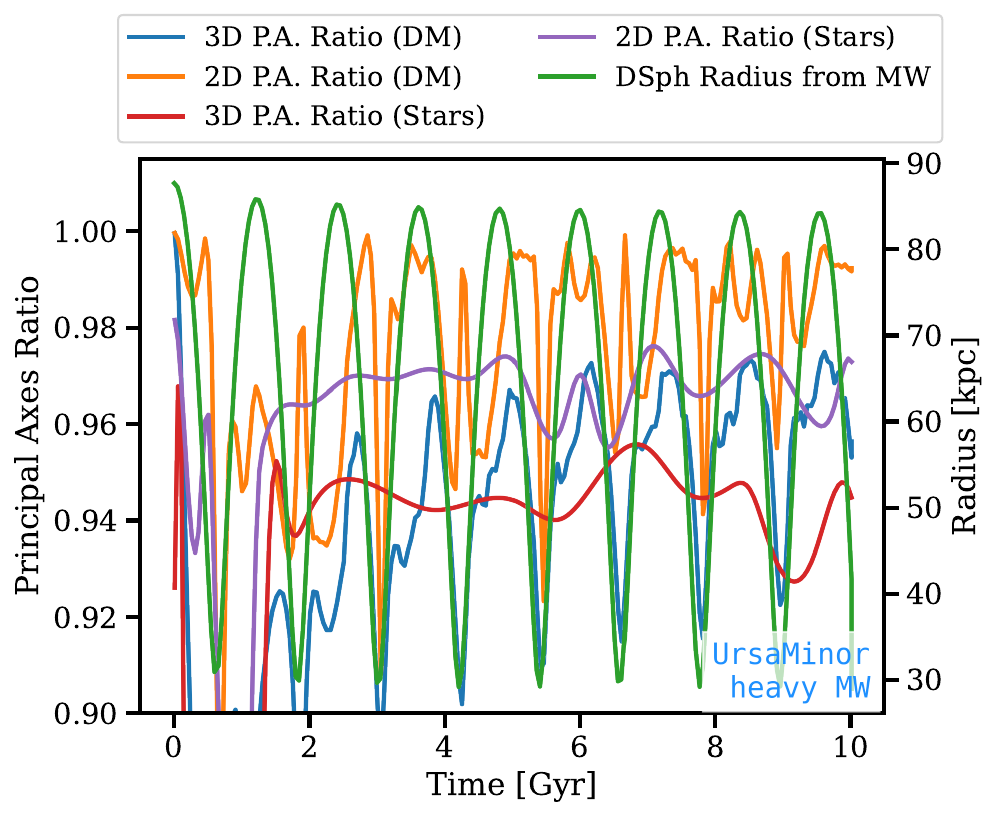}
    \caption{\textit{Top:} Ratios of the 3-D (blue) and 2-D (orange) Wolf mass estimates for dark matter tracers with the true masses in `heavy' and `light' Milky Way potentials for Ursa Minor at every snapshot of the simulation. Plotted alongside this is the distance of the dSph from the Galactic centre. We also show the modified `2-D Wolf' mass estimate (red). \textit{Middle:} Same as the top but for stellar matter. The lines show an interpolating cubic spline over the noisier results. \textit{Bottom:} Sphericities and on-the-sky ellipticities of the dark matter and stars, computed as the ratio of the minor to the major axis. For the stellar case, we plot an interpolating curve over the noisier results. Plotted alongside this is the distance of the dSph from the Galactic centre.}
    \label{fig:umi_wolf_ratio_dm}
\end{figure}

\subsubsection{Dark matter particles as tracers in Ursa Minor}
\label{wolf_dm_sec}

We begin by considering the performance of the estimator when applied to the more spatially extended population of dark matter particles as tracers. The velocity dispersion profiles of NFW haloes are expected to rise due to typically isotropic orbits in the centre found in $\Lambda$CDM simulations \citep{navarro_aquarius}, breaking the assumption of nearly flat velocity dispersion of the \citet{Wolf_2010} mass estimator. We thus proceed under the assumption that this causes a constant bias in the performance of the estimator. In the top panel of Figure \ref{fig:umi_wolf_ratio_dm}, we see an interesting oscillating effect for the two 2-D estimators and one less pronounced for the 3-D estimator, corresponding to Ursa Minor's orbit. The oscillations seem to follow a clear pattern with the orbit, with the pattern in the 2-D mass ratio computed with surface density profile fitting (orange) appearing more complex. We also note that the first few Gyr show a large inaccuracy in the estimator, which we attribute to a numerical artefact of the simulation since we suddenly plunge our dSphs into an orbit around the Milky Way potential.

To understand the origin of these oscillations, at the bottom of Figure \ref{fig:umi_wolf_ratio_dm} we show the ratio of the minor to the major axis of the dark matter halo as a function of time. We compute the principal axes using the eigenvalues of the reduced inertia tensor \citep{Bett_2007}. The blue line shows the ratio of the 3-D major and minor axes, while the orange line shows the ratio of the 2-D on-the-sky minor and major axes. It is evident that spherical symmetry is broken both in the 3-D and 2-D cases with the orbit of Ursa Minor. The dark matter halo becomes more stretched out at pericentre, leading to `peaks' in the accuracy of the 3-D mass estimator. Over time, the amplitude of the stretching becomes smaller as the halo becomes more spherical and the amplitude of the oscillations in the mass estimator accuracy appears to begin converging.

The 2-D axis ratio also demonstrates overall stretching of the dark matter halo when Ursa Minor is at pericentre, but not always. At every second pericentre, the 2-D axis ratio changes quickly from stretched-out to nearly perfect sphericity (when the dwarf is clearly deformed in 3-D). This indicates that in some instances, the 3-D major axis of the halo is aligned closely with the line-of-sight, thus causing the halo to appear spherical in 2-D. From the top panel, it can be seen that both the stretching in the plane of the sky and the times when the halo appears artificially spherical cause peaks in the estimator's accuracy.

Finally, we compare the result again with the modified 2-D Wolf estimator, which uses a directly obtained half-mass radius and mean line-of-sight velocity dispersion within this radius, shown in red in Figure \ref{fig:umi_wolf_ratio_dm}. In this case, we see that the estimator accuracy oscillates resonantly with the orbital phase, with more accurate results at apocentre and less accurate at pericentre. Only minor fluctutations in the accuracy of the modified 2-D estimator are observed at pericentre, indicating that the corresponing peaks in the 2-D and 3-D estimators stem primarily from effects of tides on the outer parts of the halo, suggesting that these effects are less likely to affect the more embedded stellar population, as we discuss in the following.

\subsubsection{Wolf estimator applied to stars in Ursa Minor}

In the middle panel of Figure \ref{fig:umi_wolf_ratio_dm}, we show the results for applying the estimator to stellar tracers in Ursa Minor. The 3-D Wolf estimate is accurate to $\sim$5-10~per~cent and displays a weak trend with the dwarf spheroidal orbit. It should be noted that the mass estimates are very noisy due to low numbers of stellar particles, which are nevertheless comparable to the largest kinematic samples available for the Milky Way’s dSphs \citep{Read2_2019}. The oscillating pattern related to the orbital phase can be seen more clearly in the 2-D mass estimator. The amplitude of the oscillations, even in the extreme case of Ursa Minor, is very small and a pattern is only present for dwarfs on particularly eccentric orbits, with eccentricity $\approx$ 0.5: Ursa Minor in the heavy potential, Draco in the light and heavy potential and Sculptor in the light potential, which is likely a result of more severe tidal shocks in these objects \citep{gnedin_shocks}. In the lower panel of Figure \ref{fig:umi_wolf_ratio_dm}, only minor changes in the projected axes are present\footnote{We note that for none of our simulated dwarfs does the 2-D axis ratio decrease beyond 0.1 (ellipticity $\epsilon = 0.1$), making it unlikely that ``stretching'' due to tides can alone explain the observed ellipticities of these dwarf spheroidal systems \citep{Iorio_2019}.}, and the 2-D mass estimator is accurate to $\sim 5$~per~cent. This is not surprising, as the stellar component is significantly more gravitationally bound than the dark matter, has a shorter dynamical time, and is, therefore, less likely to be deformed by the action of tides. It is clear from Figure \ref{fig:umi_wolf_ratio_dm}, however, that the \textit{mean} accuracy of the Wolf estimator on the stellar tracers is affected by the mass of the Milky Way. We discuss this phenomenon in the following.

\subsubsection{Light vs. Heavy Potential}\label{sec:l v h}

\begin{figure}
    \centering
    \includegraphics[width=0.95\columnwidth]{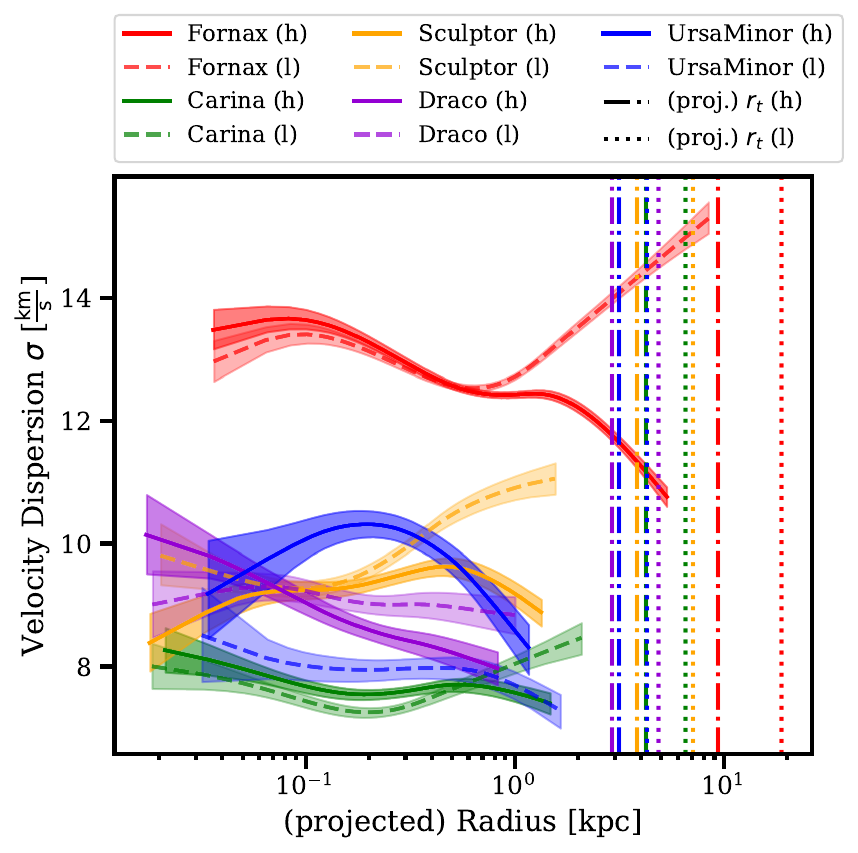}
    \caption{Best-fit line-of-sight velocity dispersion profiles obtained with {\sc pyGravSphere} for the sample of dwarfs in both `heavy' (h) and `light' (l) Milky Way potentials at their present-day snapshots. We show these instead of the dispersion profiles computed directly from the snapshot data for clarity, as all of these provide good fits. The vertical lines show corresponding tidal radii computed directly from the simulations.}
    \label{fig:vel_disp_grav}
\end{figure}

To understand what causes the systematic underestimation of mass in the `heavy' potential, in Figure \ref{fig:vel_disp_grav} we display the line-of-sight velocity dispersion profiles computed for our sample of simulated dwarfs in the two potentials. It is clear that in the `light' Milky Way potential, the line-of-sight velocity dispersion profile typically rises towards the outskirts of the galaxy (which is consistent with the kinematics of a Plummer distribution in an NFW halo in our initial conditions), while in the `heavy' Milky Way potential, the line-of-sight velocity dispersion profile tends to drop in the outskirts. Therefore, when computing the luminosity-weighted line-of-sight velocity dispersion through Equation \ref{eq:lum_w} in the `light' potential, this tends to produce a larger velocity dispersion, while the assumption of a nearly constant velocity dispersion profile in the \citet{Wolf_2010} estimator is broken. In contrast, in the `heavy' potential, the line-of-sight velocity dispersion estimate is typically lower. The use of the simplified form of the 2-D mass estimator, with velocity dispersion below the half-light radius, results in selecting parts of the velocity dispersion profile which are typically flat (aside from Draco), thus leading to lower velocity dispersion measurements and masses in the `light' potential and higher in the `heavy' potential. This trend is particularly prominent at dwarf apocentres in Figure~\ref{fig:wolf_ratios}, where the dwarfs are typically more spherical in 2-D and 3-D, thus minimizing the effects of any inaccuracies when estimating the half-light radius, or loss of equilibrium at pericentre.

We note that several sources in the literature report that rising line-of-sight velocity dispersion can be interpreted as the consequence of tides \citep{johnston_rising,readtides, munoz_tides} due to inclusion of unbound stars in the tidal tails, however in our initial conditions this is simply a consequence of modelling a Plummer stellar distribution in an NFW halo\footnote{See also \citealt{stoehr} for the velocity dispersion corresponding to the observed stellar density profiles embedded in cosmological NFW-like haloes and \citealt{penarrubia_cold} for the velocity dispersion of isotropic King profiles residing in NFW haloes, the Sculptor model of \citealt{Iorio_2019}  (who also use a Plummer model for the stellar population) also produces a similar form of the line-of-sight velocity dispersion profile.} and is particularly prominent in Fornax, Carina and Sculptor (see the Appendix). As a consequence of tides, the line-of-sight velocity dispersion profiles in our simulations tend to flatten and, in a heavy `potential', drop towards the outskirts of the galaxy when only bound stellar particles are considered. 

This opens an interesting avenue to explore in constraining the mass of the Milky Way by jointly analysing the line-of-sight velocity dispersion profiles of its satellite dwarf galaxies. However, it would be of essence to obtain data beyond the half-light radii of the dwarf satellites, into their stellar halos \citep{deason_stellar_haloes,dwarf_stellar_halos}. We caution, however, that these conclusions also depend on the internal shape of the galaxies, their stellar light distribution (for a fixed model of dark matter), and full tidal history (since orbits of many dwarfs are affected by the presence of the LMC or may have been previously tidally stripped as part of group infall). More importantly, as we show in the Supplementary Materials, the presence of unbound stars in observational samples, which are difficult to remove through standard methods like 3$\sigma$ clipping, would make it hard to distinguish the case where the velocity dispersion rises due to weak effects of tides on the stellar profile from the one where the velocity dispersion rises {\it because} of the presence of unbound stars, indicative of stronger tidal effects.

\subsubsection{Discussion}

In a recent work, \citet{errani_impulsive} presented models of adiabatic and impulsive perturbations on stellar populations residing in dark matter haloes. The former case represents the effect on the stellar populations as the dwarf approaches a pericentre of an eccentric orbit. They found that, in cuspy haloes, an impulsive perturbation results in a series of damped oscillations in the half-light radius and the velocity dispersion, which further results in oscillations in the performance of a Wolf-like mass estimator that gradually decay. This is in agreement with the pattern we saw in Section~\ref{wolf_dm_sec}, with an additional complication of minor changes in the axis ratio of the dwarfs, resulting in additional peaks in the performance of the estimator, particularly when the dwarf is viewed along the major axis. \citet{errani_impulsive} also find that in cored dark matter haloes the oscillations have a larger amplitude and are more long-lived, suggesting that it is worth exploring the performance of the estimator for individual dwarf galaxies in a similar setup to ours, but modelling the dark matter distribution with a cored profile.

Although we have simulated spherical systems, we found that tidal forces can cause deformations in the dwarfs' shape. This effect, which is the most significant at pericentre, is clear in the dark matter but can also be seen in the stars, especially for eccentric dwarf galaxy orbits. The estimator recovers true $M(r_{1/2})$ within its $1\sigma$ errors in most cases and is typically accurate within 10~per~cent, consistent with earlier findings of \citealt{errani_estimators} who applied the estimator to $N$-body counterparts of Milky Way's satellites derived from cosmological simulations. Nevertheless, our findings suggest the results of the estimator should be interpreted with care and to bear in mind the orbital properties and the current orbital phase of the subject galaxy. Additionally, it is unclear how the estimator would behave if the subject galaxy is aspherical and acted upon by tides. Our experiment with using dark matter as tracers demonstrates that for a population of tracers that extends to the tidal radius, the performance of the estimator tends to degrade, with mass typically underestimated by 15-20~per~cent until the halo stops losing significant fractions of mass, becomes more spherical, and settles towards the final bound remnant state \citep{errani_navarro}.

\begin{figure}
    \centering
    \includegraphics[width =\columnwidth]{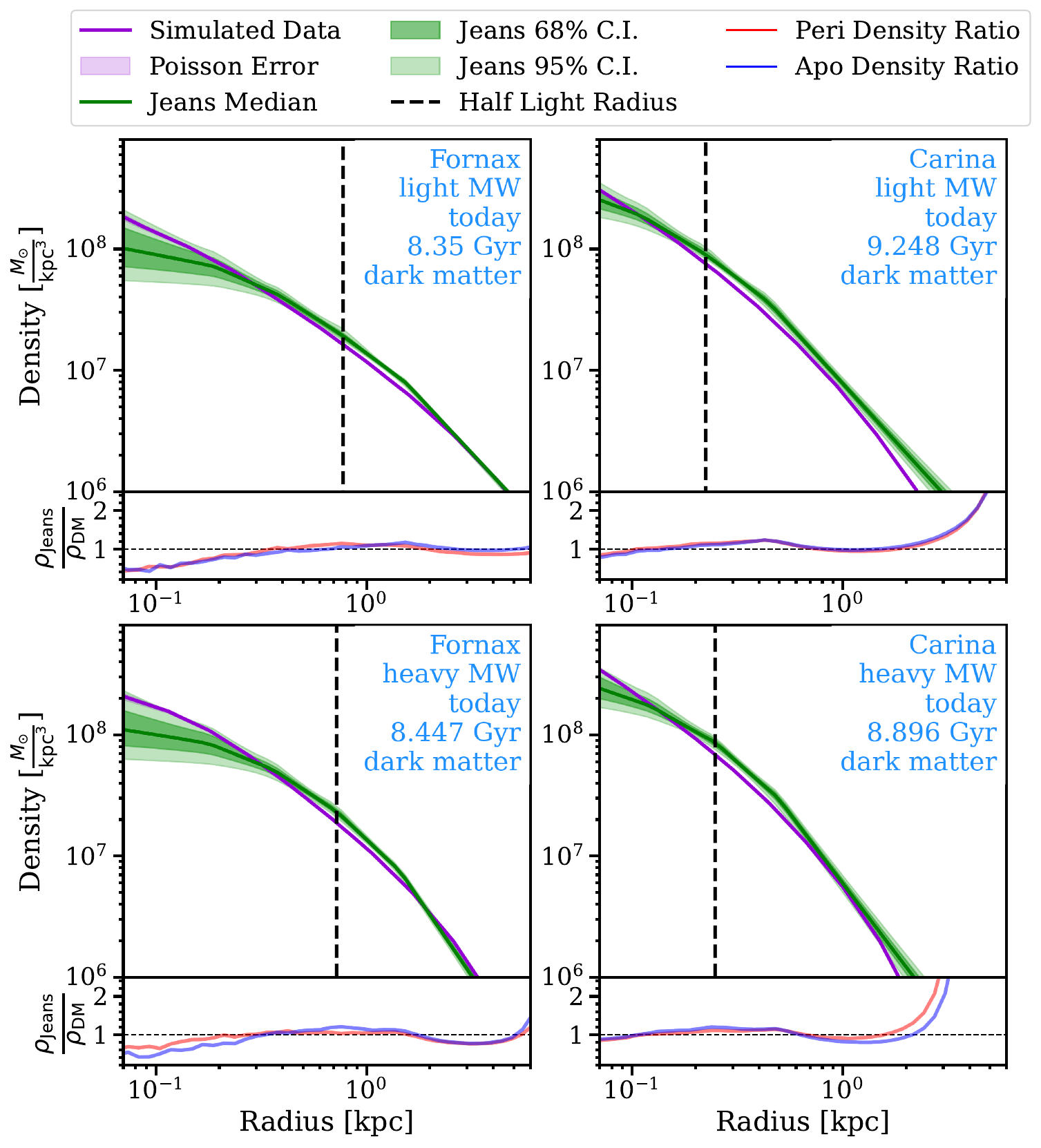}
    \caption{3-D dark matter density profiles in `heavy' and `light' Milky Way potentials of Fornax and Carina at selected simulation snapshots (closest to the position, infall time, and line-of-sight velocity of today) inferred with {\sc pyGravSphere} (green). The lighter contours show the 68~per~cent and 95~per~cent confidence limits. The `true' density profile computed directly from the snapshot is shown in purple. Below each panel are the ratios of the median recovered density profiles with the truth at apocentre and pericentre snapshots.}
    \label{fig:for_car_dens}
\end{figure}

\begin{figure*}
    \centering
    \includegraphics[width=1.8\columnwidth]{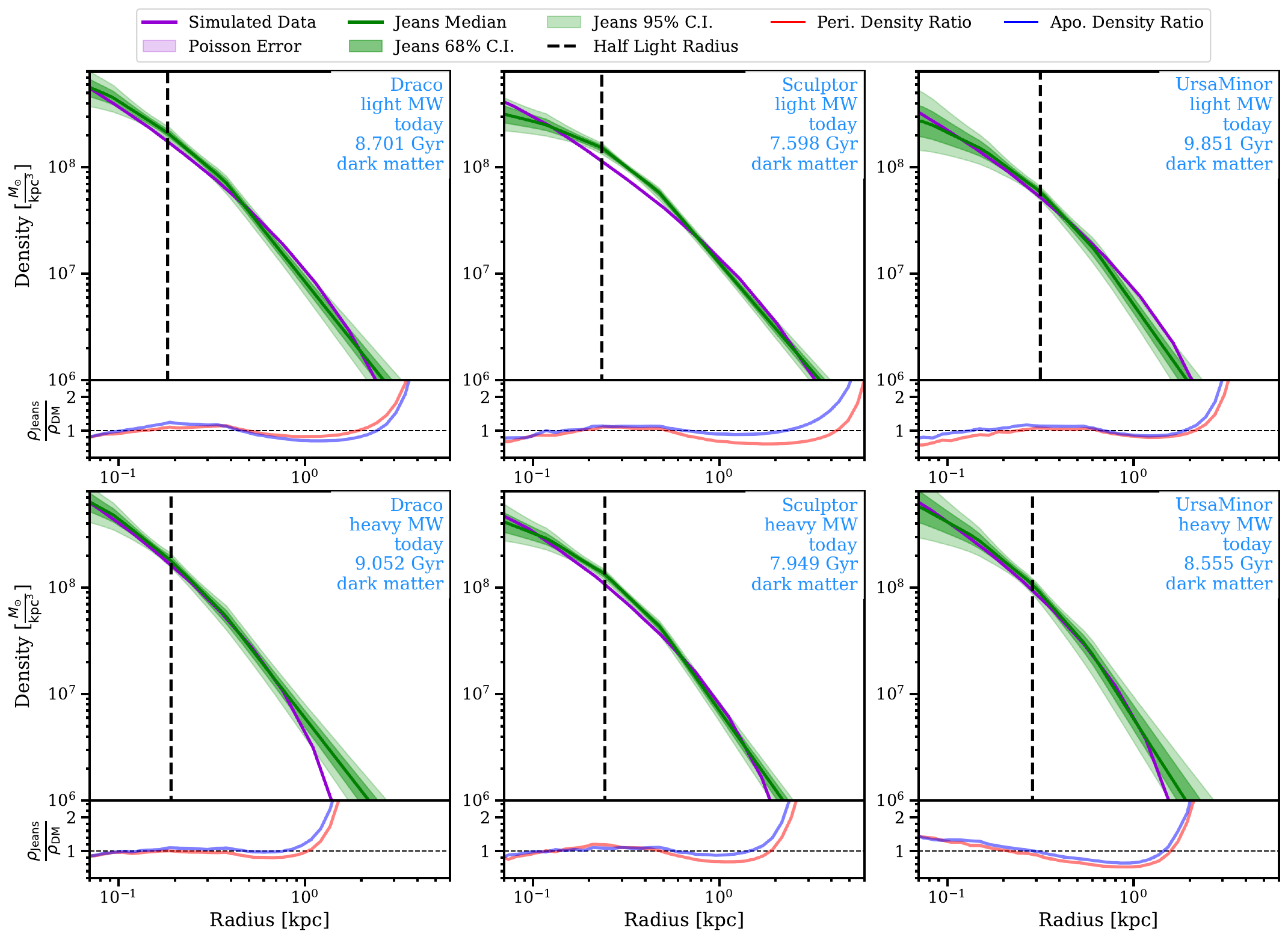}
    \caption{Same as Figure \ref{fig:for_car_dens} but for Draco, Sculptor, and Ursa Minor.}
    \label{fig:dra_scp_umi_dens}
\end{figure*}

\subsection{Results with {\sc pyGravSphere}}
\label{sec:pyg_results}
We performed the Jeans analysis on our selected sample of simulated dSphs using {\sc pyGravSphere}. This was done with 500 walkers, for 10000 steps each, while ignoring the first 5000 `burn-in' steps, which amounts to $2.5 \times 10^6$ MCMC chain steps. The priors for each parameter were uniformly distributed and are listed in Table \ref{tab:priors}.

The analysis was performed on selected snapshots from each simulation, namely, those in which the dSphs' most recent orbital phase matches that of the real observed dSphs (counting from the {\sc galpy} apocentre closest to the \citet{Rocha_2012} infall time), as well as the nearest apo- and pericentre. From the Jeans analysis, we obtain the density profiles. The median, 68 and 95~per~cent confidence intervals for the dwarfs in `heavy' and `light' potentials can be seen in Figures \ref{fig:for_car_dens} and \ref{fig:dra_scp_umi_dens} , computed from a random sample of $10^5$ MCMC chains.

\subsubsection{Effects of modelling assumptions}

Fig.~\ref{fig:for_car_dens} show the results for Carina and Fornax, which have relatively low orbital eccentricity in both potentials, with orbits close to circular. For this reason, these galaxies have bound stars extending further out than the rest of our sample, with $\sim$8~kpc for Fornax and $\sim$2~kpc for Carina. While this is further out than the samples of data used by \citet{Read2_2019}, in Supplementary Materials, we show that limiting the maximum projected radius of data produces consistent results. For Fornax in the `light' and `heavy' potential, the density is underestimated in the inner regions and overestimated near the half-light radius (black dashed line). This introduces a pattern that we will see in the remainder of our results. We note that such behaviour has been seen in the {\sc GravSphere} method before when modelling Plummer-like stellar distributions inside cuspy haloes, particularly with a high number of stellar tracers \citep{Read_2017, breaking_beta}. We have confirmed that this is not due to the permitted range of dark matter density log-slopes $\gamma$. We have tested this by modifying the priors of the broken power-law model to allow slopes as steep as $\gamma=5$. While this improved the fit to the outer profile in dwarfs that are particularly affected by tides, the fit to the inner profile did not improve, and the bias near the half-light radius remained.  Truncating the sample of stellar particles to the inner 2~kpc in the case of Fornax did not significantly improve the fit in the `light' potential in the outer regions of the halo, with the bias in the inner slope remaining. In the `heavy' potential, however, this improved the recovery of the central density (see Supplementary Materials). We also found that this effect is unlikely due to the disequilibrium of the dwarfs, as this result persists whether the dwarf is modelled at pericentre, apocentre, or in isolation. We discuss the implications of this result in Section~\ref{sec_densities}. In Appendix~\ref{sec:rbreak} we discuss in detail the origin of this bias, which is related to the inability of the broken power-law model used in {\sc pyGravSphere} to describe the shape of the dark matter profile when the stellar population is significantly embedded within the dark matter halo ($r_s /R_e \gtrapprox 4$). This is due to the model's parameterisation combined with the constraint on the ``smoothness'' of the broken power-law density profile.  

For Carina, modelling with {\sc pyGravSphere} similarly results in an overestimation of the density profile near the half-light radius. As in Fornax, this is the result of {\sc pyGravSphere} preferring a more radial anisotropy within the central kiloparsec, particularly in the `light' potential, where the inferred anisotropy is $\beta=0.1$ near the centre and $\beta=0.2$ in the outskirts of the galaxy (see Figure~\ref{fig:anis_for_car}), compared to nearly isotropic orbits in the simulated dwarf, explaining the higher densities inferred beyond the half-light radius. In the `heavy' potential, the `true' anisotropy lies within the 95$^{\rm th}$ percentiles, with a corresponding improved recovery of the density profile. 

In Fig.~\ref{fig:dra_scp_umi_dens}, we show the results for Draco, Sculptor and Ursa Minor. We note that the density profile recovery is excellent for Draco and Ursa Minor in the `heavy' potential, aside from the outermost radii (which is a consequence of our priors). Ursa Minor, for which we expect (and see) some of the strongest tidal effects (e.g. 20~per~cent loss in stellar particles and over 90~per~cent loss of dark matter particles, has its density profile constrained very well in both the `light' and `heavy' Milky Way potentials. However, due to its smaller stellar sample ($\sim30$ times smaller than that of Fornax), this introduces much larger confidence intervals through the Jeans analysis.  These smaller sample sizes then also introduce larger errors in J-factor estimations later on (see Section~\ref{section_jfactors}). 

The recovered density profile for Sculptor is somewhat reminiscent of the results for Carina, also with a preference for more radial velocity anisotropy (see Figure~\ref{fig:anis_dra_scp_umi}). We note that Fornax, Sculptor and Carina also have the largest kinematic samples of the dwarfs we consider, so it is expected that the biases inherent in the modelling would become more prominent \citep{breaking_beta}. Moreover, it is clear from Figure~\ref{fig:vel_disp_grav} that Fornax, Sculptor and Carina are the three galaxies in our sample that still exhibit a Plummer-characteristic rise of the velocity dispersion in the outer regions (see Section~\ref{sec_theory} of the Appendix), while Draco and Ursa Minor are nearly flat or declining. We suspect that this rise in the velocity dispersion profile, the peak of which approximates well the scale radius of the halo (see Appendix A), is, in part, what leads {\sc pyGravSphere} to infer higher densities near the half-light radius, which lowers the inferred central density. We discuss this further in the Supplementary Materials, where we also show that for rising 4-th velocity moments, the method of \citet{read_draco} tends to over- or underestimate the virial shape parameters depending on the spatial extent of the data, which affects the constraint on the outer density profile and, through the parameterisation of the model, the inferred densities in the inner regions.

\subsubsection{Effects of tides}

We have additionally applied the Jeans analysis to the nearest pericentre and apocentre snapshots of the dwarfs. The comparison of the accuracy of the recovered profiles is shown in the subpanels of Figures~\ref{fig:for_car_dens} and \ref{fig:dra_scp_umi_dens}. Overall, we see that the quality of the density profile recovery is consistent with that of the present-day snapshots. Our analysis with the \citet{Wolf_2010} mass estimator, even in the extreme case of Ursa Minor, also predicts variation of below the 10~per~cent level for our sample of dwarf galaxies. The main discrepancy is in the outskirts of the dwarfs, where the density inferred at apocentre is overestimated compared to the pericentre. This may seem counterintuitive, but the majority of the mass loss occurs after the pericentric passage is complete, such that the outer density slope is higher at the next apocentre than at pericentre \citep{penarrubiatides}. Since our priors assume density slopes $\gamma$ between 0 and 3, the inferred density is, therefore, more inaccurate at apocentre (if the apocentre snapshot is after the pericentre). Note that this trend is reversed for Carina in the heavy potential, where the pericentre snapshot is after the apocentre one. 

Ursa Minor in the `heavy' potential shows systematic overestimation of the central density at both the peri- and apocentre snapshots. These snapshots occur after the present-day snapshot, and we find that the central line-of-sight velocity dispersion at these times is increased, transforming the velocity dispersion profile from a fairly flat to a declining one. This is because the stellar component becomes more strongly deformed and heated by tidal forces (see bottom panel of Figure~\ref{fig:umi_wolf_ratio_dm}), although we do not see signs of unbound stars in terms of the line-of-sight velocity-distance distribution \citep{klimentowski2}, with all stars passing the $3\sigma$ clipping test.

The bias in the recovered profiles is consistent with those inferred from the {\sc Auriga} simulations by \citet{wenting} (i.e. an underestimation of density in the central regions and overestimation in the outer regions); however, these authors attribute this to contraction of Sagittarius-like systems due to tides. We note that their modelling also fixes the maximum value of the outer halo slope to 3, which would explain the overestimated density in the outskirts of their dwarfs. The reason for their underestimation of the inner density is less clear but could be explained if their priors do not allow steeper inner slopes than $\gamma=1$. While this result may be due to the halo contraction induced by tides, in our analysis with {\sc pyGravSphere}, this inner density underestimation occurs independently of the orbital stage or distance from the Milky Way.

Our results may seem at odds with those of \citet{Genina_2020}, who tested {\sc pyGravSphere} on Fornax-like systems in the APOSTLE cosmological simulations and generally found a minor underestimation of the density with {\sc pyGravSphere} at all radii; however, the systems that they consider depart from sphericity, which drives overestimation or underestimation of the enclosed mass, have smaller samples of stellar particles ($<2500$, on par with our models of Draco and Ursa Minor), with the spatial extent of the data limited to within 2 half-light radii, and lack the spatial resolution to analyse densities recovered below $\approx 300$~pc. The overestimation of the density at large radii due to the dark matter density log-slopes $\gamma$ being limited to 3 is in agreement with their results.

Previous works of \citet{battagliaFornaxtides} and \citet{Iorio_2019} are the most similar in nature to ours, considering Fornax and Sculptor dwarf galaxies, respectively, tidally stripped in the Milky Way potential and on orbits constrained by the measured proper motions. Both works additionally apply Jeans dynamical modelling to their simulated dwarfs at the present-day snapshots. While there are some differences in the modelling choices, in particular, the details of the Milky Way potential and the dark matter distribution of the dwarf galaxies, both of these previous studies agree with our conclusions that for Sculptor and Fornax the assumption of dynamical equilibrium still holds. The work of \citet{Iorio_2019} in particular has explored the most eccentric orbits permitted by proper motion errors and have taken a more observational approach to selecting the kinematic sample in their simulations for the dynamical modelling (based on selection function for observational data as opposed to considering all bound particles), they still conclude that tidal effects are not significant on the inferred density and mass distributions. Similar conclusions were reached in \citet{read_draco}, who applied the {\sc GravSphere} model to mock data generated from simulations of the properties and orbit of Draco. They found that {\sc GravSphere} was able to faithfully recover the density profile of Draco after 10~Gyr of evolution in the Milky Way potential. As such, the results presented here are in good agreement with previous work, despite differences in modelling choices.

The work of \citet{munoz_tides} models the orbital evolution of Carina dwarf spheroidal and finds a significant impact of tides, although they do not apply dynamical modelling to their modelled satellites. The difference from our results likely stems from the fact that they model the dark matter distribution of Carina with $10^5$ particles and using a cored \citet{plummer} model. Both of these choices would result in enhanced physical and artificial tidal disruption of their modelled satellites \citep{van_den_Bosch_2018, errani_cores}.

To conclude, we find that for the sample of 5 initially spherical dwarfs in 2 Milky Way potentials considered in this work, the assumption of dynamical equilibrium is a reasonable one; instead, the inaccuracies we found stem from the parameterization of the dynamical models. As we only considered bound tracers in our analysis, we suggest that the presence of unbound stars in kinematic samples pose a greater risk to the accuracy of mass distributions recovered with spherical Jeans models for these dwarfs than the perturbations to the dynamical equilibrium \citep{klimentowski, lokas_tides}.

\subsection{Dark matter central densities}
\label{sec_densities}
\begin{figure}
    \centering
    \includegraphics[width=0.95\columnwidth]{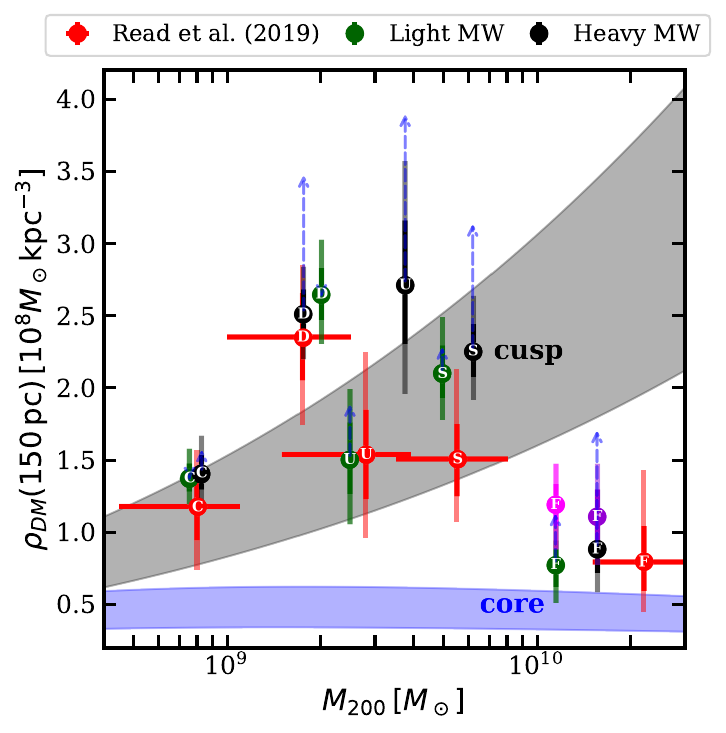}
    \caption{Dark matter densities at 150~pc of the simulated dSphs obtained with {\sc pyGravSphere} with their 68~per~cent (solid) and 95~per~cent (fainter) confidence intervals, in both the `light' and `heavy' Milky Way potentials. On the horizontal axis are the initial masses of the simulated NFW haloes. The blue dashed arrows point to the inner densities calculated from the initial snapshots of the simulations. C = Carina, D = Draco, U = Ursa Minor, S = Sculptor, F = Fornax. The dashed grey contour is the densities expected from the mass-concentration relation in $\Lambda$CDM \citep{Ludlow_2016} and the blue contour is the density expected if a core forms below the half-light radius \citep{kravtsov}. Red symbols show the densities computed by \citet{Read2_2019} and pre-infall halo masses from \citet{Read_2019}. The pink and purple data points are the same inner densities for Fornax in the `light' and `heavy' Milky Way potentials, but obtained assuming a \citet{Zhao_1997} profile and virial shape parameters with {\sc pyGravSphere}.}
    \label{fig:comb_dens}
\end{figure}

In Figure \ref{fig:comb_dens}, we display the densities at 150 pc computed with {\sc pyGravSphere} as in \citet{Read2_2019}, but for our sample of simulated dwarfs. The grey band displays the prediction of the mass-concentration relation in $\Lambda$CDM \citep{Ludlow_2016}, while the blue band displays the prediction if all galaxies formed dark matter cores within their half-light radius \citep{kravtsov}. The red points with error bars show results from \citet{Read2_2019}, while the black and green points show the results for our simulated dwarfs in the `heavy' and `light' potential, respectively. There is a remarkable agreement between the results of \citet{Read2_2019} and our models of Carina and Draco. Results for Ursa Minor agree within $1\sigma$ in the light potential, with a 2$\sigma$ discrepancy in the `heavy' potential. Sculptor is consistent with the result of \citet{Read2_2019} within $2\sigma$. Fornax in the `heavy' and the `light' potential is consistent with the result of \citet{Read2_2019} within $1\sigma$.

As we have shown in Fig.~\ref{fig:c_mass_comb}, some of our models in the initial conditions are outliers in the mass-concentration relation, in particular the two models for Draco and the `heavy' potential models for Ursa Minor and Sculptor. Their initial inner densities are shown with blue arrows in Figure~\ref{fig:comb_dens}. Through mass loss due to tidal stripping, their present-day inner densities \textit{appear} in good agreement with the mass-concentration relation, however. In the future, with more realistic models of the Milky Way potential and the LMC, it would be informative to explore whether NFW haloes are compatible with the present-day densities of Milky Way's satellites when tidal stripping is taken into account. Our results suggest that insignificant tidal stripping (i.e. a `light' Milky Way potential) can put the observed kinematics of Milky Way's dwarfs in agreement with $\Lambda$CDM, but would still require Draco and Fornax to be outliers in the mass-concentration relation (assuming 0.1~dex scatter in the relation at the low-mass end). It is also unclear how this result fits into the `tidal stirring' scenario of dwarf spheroidal formation from initially disky galaxies \citep{tidal_stirring}, as it seems unlikely that Milky Way's tides can affect the stellar populations of the five dwarfs considered in this work significantly \citep{ficintio_fornax_strip}, especially if the infall time of these dwarfs is late \citep{battaglia_edr3}. On the other hand, the preferred stripping of stars on prograde tangentially anisotropic orbits may still reconcile this apparent inconsistency \citep{readtides,stirring_prograde}, but will require further tests with simulations to confirm this. Sudden loss of gas due to ram pressure and subsequent tidal effects may also alleviate the tension \citep{hammer, wang_hydro_mond}. 

We additionally note that, in part, the agreement we see here results from {\sc pyGravSphere}'s tendency to underestimate the central densities. This is particularly apparent with Fornax in the `heavy' and `light' potential, with an additional reduction of the central density due to tides \citep{genina_fornax}. We have explored modelling our simulated systems with the \citet{Zhao_1997} dark matter density model available in {\sc pyGravSphere} and found a more faithful recovery of the inner density profile (purple and pink points in Fig.~\ref{fig:comb_dens}). This is partly because the \citet{Zhao_1997} parameterisation is less flexible in its ability to model small dark matter cores or tidal truncation, thus forcing the density profile towards a more cuspy solution, which in our case is the correct one, but it may not be in nature. Despite this apparent success of the \citet{Zhao_1997} model in this case, we would argue for more non-parametric approaches to dynamical modelling, which will more likely return accurate solutions in the future as distances and proper motions become more available for individual stars in nearby dwarf spheroidals.

\subsection{J-Factor Results}
\label{section_jfactors}
\begin{figure}
    \centering
    \includegraphics[width=\columnwidth]{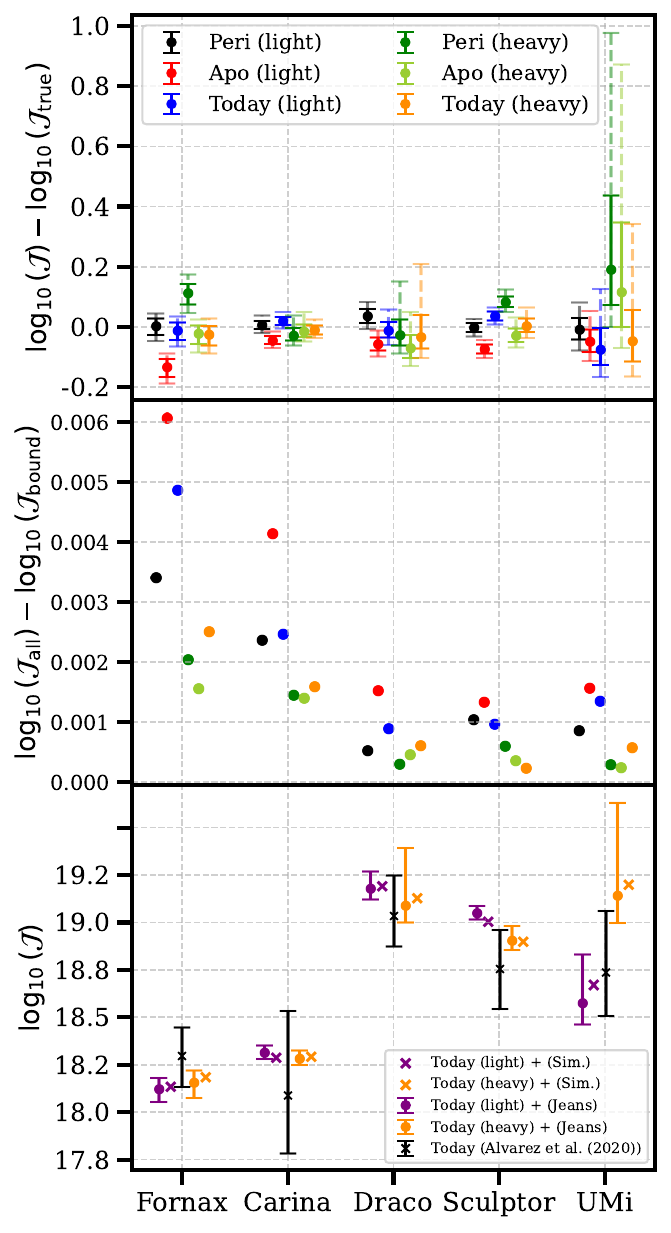}
    \caption{\textit{Top:} logarithm of the ratio of J-factors computed using best-fitting dark matter densities derived with {\sc pyGravSphere} to the `true' J-factors obtained directly from the simulations at present-day, peri and apocentre snapshots. The error bars show the median, as well as the 68 and 95~per~cent confidence limits. The J-factors are computed within the central 0.5~degrees of the dwarf. \textit{Middle:} difference in the `true' J-factors calculated using only the bound dark matter particles, compared to all particles, bound and unbound. \textit{Bottom:} absolute values of the J-factors computed in this work, compared to those estimated using the same set of models in {\sc GravSphere} using observational data in \citet{calore_jfac}. }
    \label{fig:comb_jf_errs}
\end{figure}

We calculate the J-factors of a sample of $\sim3 \times 10^5$ MCMC chain steps for each dSph directly from the Jeans analysis results, as well as the `true' J-factors of the corresponding snapshots. The top panel of Figure \ref{fig:comb_jf_errs} summarises the results for the dSphs at present-day snapshots and when the dSphs were closest to their peri- and apocentres in the `light' and `heavy' Milky Way potential.

As a general trend, we find that the J-factors are typically underestimated for our simulated systems, which we attribute to the lower estimated central densities. Exceptions to this are Carina and Sculptor in the light potential, where there is a significant overestimation of the density within the central 1~kpc. The discrepancy in the outer profiles, which is primarily caused by our priors, would correspond to an insignificant contribution to the J-factors since regions below $r_{\rm max}\approx2r_s$ dominate the contribution to the J-factor in NFW haloes \citep{springel_jfacs}. At apocentre, the J-factors are likely to be underestimated in both the `light' and `heavy' potential. This is likely the result of the dwarfs orienting along the line of sight at apocentre that increases the `true' J-factor. This is evident in the middle panel of Fig.~\ref{fig:comb_jf_errs}, where we show the comparison between the `true' J-factors calculated only for the bound particles and those computed using all particles within the solid angle. Although the difference is rather small due to significantly lower densities of the stripped dark matter, it is clear that in the `light' potential, the stripped debris tends to align with the line of sight at apocentre. In the `heavy' potential, this difference is less significant as the dark matter in the dwarfs has already undergone significant stripping, and they lose progressively smaller fractions of their mass.

For present-day snapshots, the true value is typically contained within the 95~per~cent confidence intervals. The difference between the inferred J-factors and the `true' ones is generally small but not negligible. It is also clear that in some cases, the errors on the J-factors are underestimated. While our J-factors computed with {\sc pyGravSphere} are in good agreement with those of \citet{calore_jfac} (bottom panel of Fig.~\ref{fig:comb_jf_errs}), who also used the {\sc GravSphere} technique, our results suggest that the `true' error on the J-factors should be at least of order $\sigma_J \approx 0.1$, based on our modelling of tidally stripped and initially spherically-symmetric systems with {\sc pyGravSphere}. 

\section{Conclusions}
\label{conclusions_section}

The spherical Jeans equation remains a popular tool to model the dynamics of dwarf galaxies. Dark matter density distributions are sensitive probes of the nature of dark matter. They are a crucial quantity in determining the expected flux of the by-products of dark matter annihilation and decay, which can be detected with space and ground-based instruments, thus putting a constraint on the particle-physics nature of dark matter. Dwarf spheroidal satellites of the Milky Way represent particularly interesting objects in which to study dark matter due to their proximity and sensitivity to dark matter and baryonic effects; however, as satellites of the Milky Way, they are subject to Milky Way's tidal forces, which can break both the assumption of spherical symmetry and dynamical equilibrium underlying the spherical Jeans equation.

In this work, we tested the performance of the spherical Jeans equation on tidally-perturbed dwarf galaxies by producing a suite of tailored simulations of Fornax, Carina, Sculptor, Draco and Ursa Minor dwarf spheroidals in `light' and `heavy' Milky Way potentials and applying the non-parametric Jeans analysis tool {\sc pyGravSphere} to the simulated dwarfs. Our findings can be summarised as follows:

\textit{i)} NFW haloes with stellar distributions that are well-fit by a Plummer profile are able to match the projected half-light radii and velocity dispersions within the half-light observed in Fornax, Carina, Sculptor, Draco and Ursa Minor at the present day. For the `light' potential, the NFW haloes are consistent with the mass-concentration relation in $\Lambda$CDM, however in the `heavy' potential, Draco and Ursa Minor would need to be outliers from the mass-concentration relation. There is additionally a discrepancy for Fornax in the `light' potential.

\textit{ii)} Plummer profiles in NFW haloes lead to peculiar shapes of the line-of-sight velocity dispersion, with profiles that typically fall towards the half-light radius and then rise. We, therefore, note that rising line-of-sight velocity dispersions in the outskirts of dwarfs are not necessarily a signature of tides but rather a natural outcome of Plummer distributions embedded in spherical NFW haloes.

\textit{iii)} In the `light' Milky Way potential, the part of the line-of-sight velocity profile that rises is generally maintained for our sample of dwarfs, while in the `heavy' potential, the profile flattens or even steepens in the outskirts when only the bound stellar particles are considered, suggesting a possibility of inferring the mass of the Milky Way based on the velocity dispersion of its satellites. This prospect is, however, complicated by the need to obtain kinematic data samples that are free from contamination by unbound stars.

\textit{iv)} When the \citet{Wolf_2010} mass estimator is applied to these systems, the accuracy of the estimator depends on how the velocity-dispersion is derived. If the luminosity-weighted line-of-sight velocity dispersion is used, the accuracy of the estimator is affected by the rising or falling shape of the velocity dispersion profile in the `light' and `heavy' potential, respectively. The inferred masses are larger when the velocity dispersion profile is rising and lower when they fall. In some cases, the accuracy of the estimator can be improved by instead using the velocity dispersion within the half-light radius in the mass estimator. Nevertheless, we found both forms of the estimator to be accurate to better than 10~per~cent for our simulated dwarf spheroidals at the present day (see also \citealt{errani_estimators}), though we emphasise that we consider initially spherical systems.

\textit{v)} We applied the \citet{Wolf_2010} mass estimator to both the dark matter and stellar component over the entire orbital history of the dwarfs. For dark matter, we observed a clear correlation between the accuracy of the estimator and the orbital phase. This is partially related to the asphericity of the dark matter halo induced by tides; however, the ratio of the minor to major axes does not typically drop below 0.9 and is, therefore, unlikely to dominate the effect. For the stellar component, there is no such clear correlation as the stars are significantly more bound and ``shielded'' by the dark matter. Nevertheless, for dwarfs on particularly eccentric orbits, we find a correlation between estimator accuracy and orbital phase also in the stars. This is likely due to the more enhanced tidal shocks in these objects \citep{gnedin_shocks}. Our results suggest that the orbital phase of the satellite should be taken into account when interpreting the results of the mass estimator, particularly in cases where the stellar population is extended and likely approaches the size of the dark matter component, such as ultra-diffuse galaxies like Crater II and Antlia II \citep{craterii,antliaii}.

\textit{vi)} We applied the non-parametric Jeans code {\sc pyGravSphere} \citep{Read_2017,Genina_2020} to our sample of simulated dwarfs and found that the severity of tidal effects had no significant effect on the quality of the density profile recovery when analysing only bound stellar tracers, aside from the ability to model the outer parts of the dark matter profile, which is partially related to the density slope priors that we set. We conclude that, for the sample of dwarfs considered in this work, the presence of unbound stars in kinematic samples likely poses a greater challenge to Jeans models than the dynamical perturbations due to tides themselves. 

We also found a typical underestimation of the central density \citep{Read_2017,breaking_beta} which arises due to the inability of the broken power-law model to fit the dark matter halo profile, combined with the ``smoothness'' constraint on the broken power-law density profiles. The tendency for the virial shape parameters to be either under- or over-estimated if the true extent of the galaxy is not fully known (see the Supplementary materials and the discussion in \citealt{read_draco}) can worsen this issue. We thus suggest that dynamical models with {\sc pyGravSphere} should either keep the power-law break radii as free parameters of the model, or to increase the number of bins until the convergence in the obtained result is reached (though as \citet{Read_2017} point out, an increase in the number of bins should also be motivated by the quality and quantity of the data). Additionally, we suggest that the virial shape parameters should be estimated only for the available extent of observational data \citep{richardson} and used as lower limits on the virial shape parameters of {\sc pyGravSphere} models.

\textit{vii)} Comparing the inferred densities at 150~pc and the pre-infall $M_{200}$ of the dwarfs to the results of \citet{Read2_2019}, we found excellent agreement, particularly for Carina and Draco and generally dwarfs in the `light' Milky Way potential. Fornax in the `heavy' and `light' potentials was consistent with the measurement of \citet{Read2_2019}, which was previously taken as a sign of reduced central density in this galaxy due to repeated heating of the dark matter component by supernovae feedback, whereas we do not take baryonic effects into account in this work. The low density of Fornax in our simulations is instead a consequence of it being an outlier in the mass-concentration relation, which is more significant in the `light' potential, combined with the tendency to underestimate the central density by {\sc pyGravSphere}. 

\textit{viii)} We have computed astrophysical J-factors within 0.5 degrees using the inferred density profiles with {\sc pyGravSphere} and compared those to the true values, obtained directly from the simulations. We found a general tendency to underestimate the J-factor, and, in many cases, the error bars for the J-factors were underestimated, although the absolute discrepancy from the true value is typically small. Assuming the effects of asphericity are insignificant, we suggest that the error on the inferred J-factors with {\sc pyGravSphere} should be at least $\sigma_J = 0.1$~dex to encompass the true value of the J-factor. 

\textit{ix)} We have made the snapshots of simulations analysed in this work publicly available and hope that these can serve as mock datasets to test alternative dynamical modelling techniques in the future. It would be straightforward to modify these mocks to e.g. include a contaminating sample of background Milky Way stars, model stellar binaries and explore different observational sample selections. 
\newline

To conclude, while our simulations may not fully represent the orbital and tidal evolution of the five dwarfs, they likely cap the most and least extreme scenarios and, for those, we found that tides are unlikely to affect the quality of density profiles inferred with Jeans analysis if the presence of unbound stars has been accounted for, with asphericity being likely a more important effect for these objects. We have seen some minor effects of tides on the accuracy of mass inferred for the Ursa Minor dwarf spheroidal. This suggests that tailored simulations like ours would be an invaluable tool to interpret the results of mass estimators and Jeans analysis for dwarfs on lower-pericentre and higher-eccentricity orbits than the sample considered here. 

It would, of course, be interesting to confirm our findings with simulations that account for the influence of the Large Magellanic Cloud on dwarf spheroidal orbits as well as the Milky Way's mass evolution \citep{buist_helmi,vasiliev_time_varying}. Our models of the dwarf spheroidals are also not fully realistic, as they do not take into account the possible initial disky configuration of the dwarfs \citep{tidal_stirring}, their present-day asphericity \citep{schelkanova}, rotation \citep{dwarf_rotation}, multiple kinematically-distinct stellar populations \citep{tolstoy,battaglia06,pace_ursaminor} or the influence of gas and star formation on their potential wells \citep{pontzen_governato}. We have also not considered initially cored dark matter density distributions \citep{stirring_density_profile}, and thus, it is unclear whether {\sc pyGravSphere} would recover those accurately in the presence of tidal effects.

Tidal forces are not the only way to violate the equilibrium assumption of the Jeans equation; recent mergers can lead to similar, potentially more long-lasting effects. With evidence of `shells' in e.g. Fornax dwarf spheroidal \citep{fornax_shells, fornax_merger} and possible merger origin kinematically hot extended stellar populations \citep{genina_pops}, it is also important to evaluate the past merger history of Local Group dwarfs and whether those lead to phase-mixed stellar populations that can be faithfully modelled with the Jeans equation at the present day.

\section*{Acknowledgements}
KT and AG would like to thank Sherry Suyu and Volker Springel, without whom this work would not have been possible and Thorsten Naab for continued support. Additionally, the authors would like to thank Simon White and Klaus Dolag for helpful discussions and Justin Read for providing useful feedback on the manuscript and explaining the effect of the ``smoothness'' constraint on the broken power-law model. We also thank the anonymous referee for insightful comments. This research was carried out using the High-Performance Computing resources of the FREYA cluster at the Max Planck Computing and Data Facility (MPCDF, https://www.mpcdf.mpg.de) in Garching, operated by the Max Planck Society (MPG).

\section*{Data Availability}

The simulation snapshots analysed in this work and {\sc pyGravSphere} input files can be found at: \url{https://keeper.mpdl.mpg.de/d/04376cbd45414bdab4a1/}

Other data in this article will be made available upon reasonable request to the corresponding author.



\bibliographystyle{mnras}
\bibliography{example} 



\FloatBarrier
\appendix
\renewcommand\thefigure{A\arabic{figure}}    
\setcounter{figure}{0} 

\label{Appendix}

\section{Line-of-sight velocity dispersions of Plummer profiles in NFW haloes}
\label{sec_theory}
\begin{figure}
    \centering
    \includegraphics[width=0.9\columnwidth]{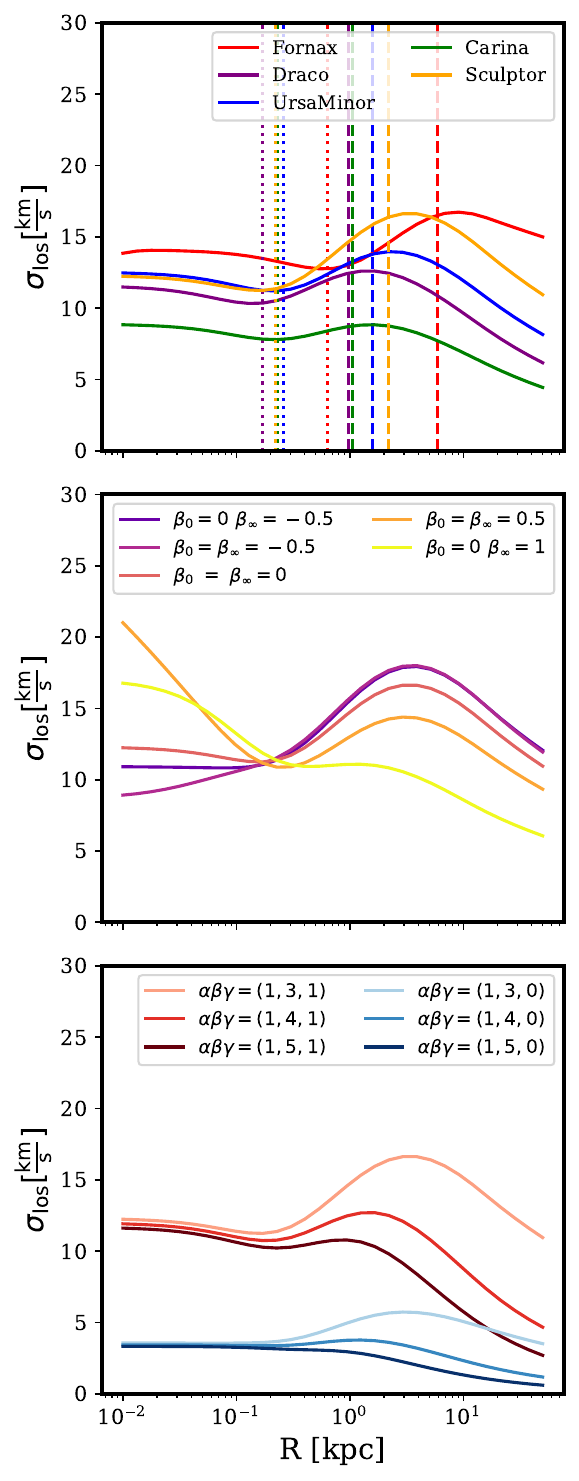}
    \caption{\textit{Top:} stellar line-of-sight velocity dispersion profiles of the exact solutions of the Jeans equations for the `heavy' Milky Way potential Plummer and NFW parameters of our sample of dwarfs. Note that these will be slightly different from the velocity dispersion profiles in our initial conditions due to the exponential truncation of the profiles and anisotropy that develops due to the distribution function disfavouring pure isotropic solutions \citep{almeida_cores_in_cusps}. The dotted lines show the locations of the Plummer half-light radius, while the dashed lines show the locations of the NFW scale radius, $r_s$. \textit{Middle:} varying the velocity anisotropy for the Sculptor `heavy' model, with the radius $r_a$ set to the half-light radius. \textit{Bottom:} for the same values of $r_s$ and $\rho_s$, varying the inner and outer slopes of the Sculptor `heavy' model dark matter profile in the \citet{Zhao_1997} formulation, while the velocity anisotropy $\beta$ is set to 0 (isotropic). For cored haloes ($\gamma=0$), the scale radius is set to 1~kpc.}
    \label{fig:disp-theory}
\end{figure}

In the top panel of Figure~\ref{fig:disp-theory}, we show the shape of the line-of-sight velocity dispersion profiles of Plummer stellar distributions residing in NFW haloes, assuming full isotropy. The profiles were obtained by solving the spherical Jeans equation. We note that a fully isotropic solution is not permitted by the condition of the non-negativity of the distribution function \citep{almeida_cores_in_cusps}; however, since we allow our generated models to settle in equilibrium, they acquire some anisotropy (see Figures~\ref{fig:anis_for_car} and \ref{fig:anis_dra_scp_umi}). The dotted lines show the half-light radii of our simulated dwarfs in the initial conditions, which align well with a `dip' in the velocity dispersion profile. The dashed lines show the location of the NFW scale radius, $r_s$, which slightly precedes the peak of the `bump' in the velocity dispersion profile. 

The middle panel shows that the radial anisotropy in the outskirts tends to lower the velocity dispersion of the outer `bump', while radial anisotropy in the inner region increases the velocity dispersion in the centre. Tangential anisotropy in the outskirts causes the velocity dispersion in the outer regions to rise. The `bump' beyond the half-light radius is preserved with the change in the anisotropy profile but can be wiped away by tidal forces, or may not be fully captured depending on the extent of the observational data. Variations in the mass profile can, of course, also lead to changes in the shape of the profile, with the classic mass-anisotropy degeneracy, as we demonstrate in the bottom panel for the case of cored and cuspy haloes, for which we change the steepness of the outer slope of the dark matter profile. In particular, outer slopes $\gamma=4$ or 5, significantly reduce the amplitude of the outer `bump'. This likely explains why our `heavy' models do not exhibit a strong `bump' in the line-of-sight velocity dispersion, despite having tangential anisotropy in the outskirts \citep{readtides, penarrubiatides}.

\renewcommand\thefigure{B\arabic{figure}}    
\setcounter{figure}{0}

\begin{figure}
    \centering
    \includegraphics[width = \columnwidth]{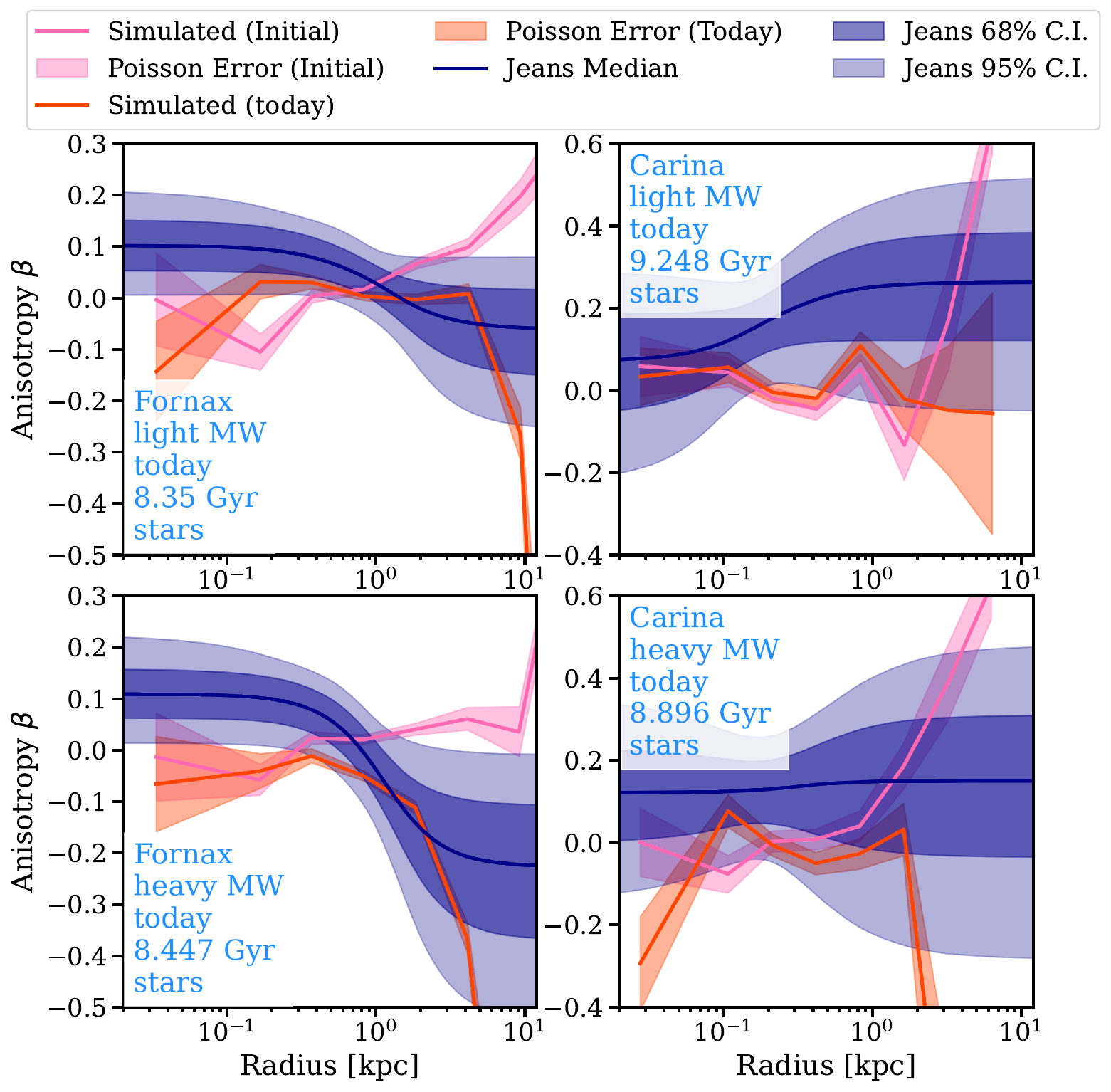}
    \caption{Stellar anisotropy profiles for Fornax and Carina in both the `light' and `heavy' potential at the present-day snapshot. The purple profiles are the anisotropy profiles recovered by {\sc pyGravSphere} with 68~per~cent and 95~per~cent confidence intervals. The pink curves are the profiles calculated from the initial simulation snapshot. The red curves are the same as the pink but calculated from the present-day snapshots using all bound particles. We note that this is not the same radial extent as that of the projected data used in {\sc pyGravSphere} modelling (Figure~\ref{fig:vel_disp_grav}).}
    \label{fig:anis_for_car}
\end{figure}

\begin{figure*}
    \centering
    \includegraphics[width =1.8\columnwidth]{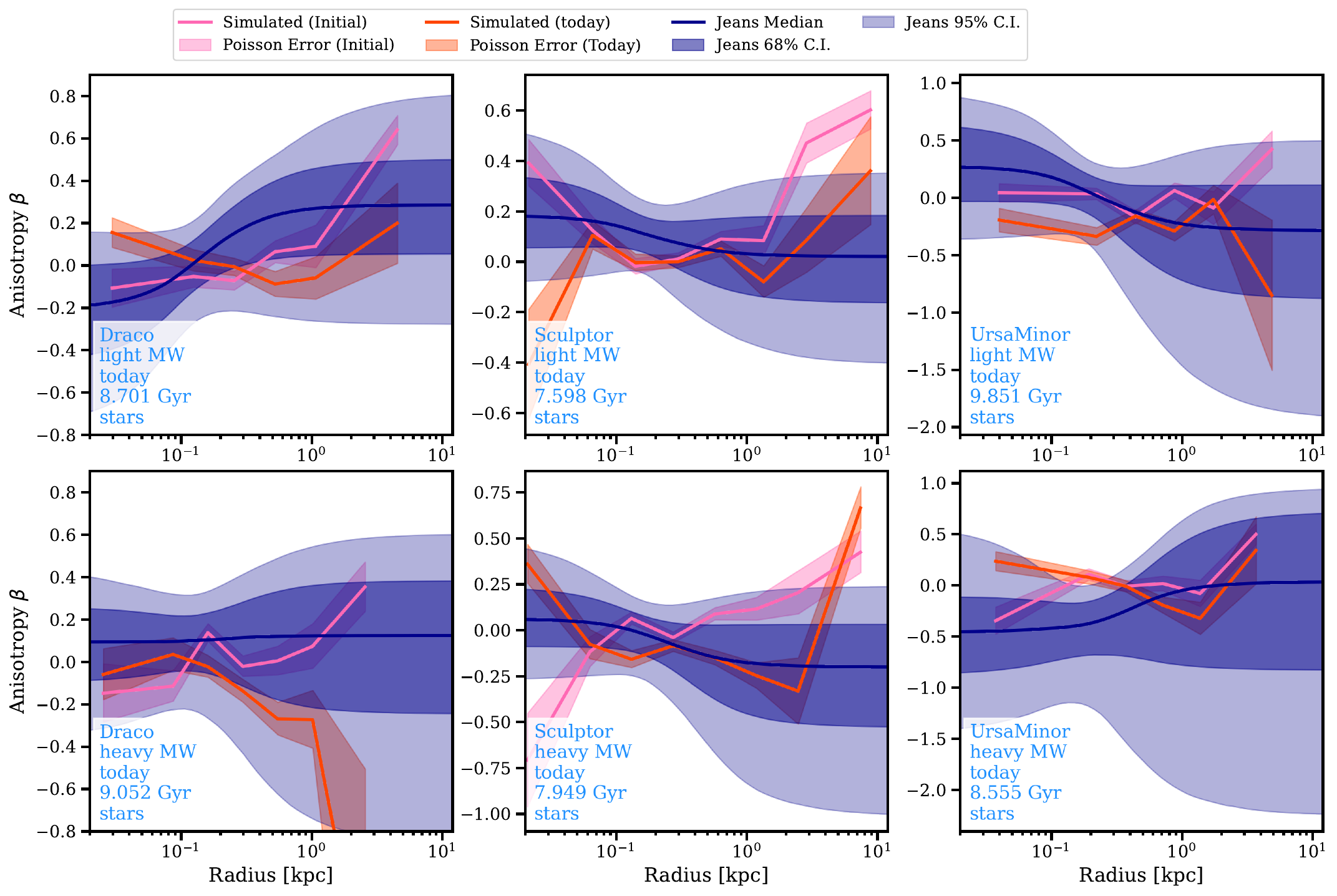}
    \caption{Same as Figure \ref{fig:anis_for_car} but for Draco, Sculptor, and Ursa Minor.}
    \label{fig:anis_dra_scp_umi}
\end{figure*}

\section{Velocity anisotropy recovery}
\label{anisotropy_recovery}

In Figures~\ref{fig:anis_for_car} and \ref{fig:anis_dra_scp_umi}, we show the quality of the constrained velocity anisotropies by {\sc pyGravSphere} (purple contours). The pink line shows the anisotropy profile in the initial conditions (after the system was left to settle for 5~Gyr) and the red line shows the anisotropy profile at the present day. We note that our simulated systems are not fully isotropic in the initial conditions but typically acquire mildly radial or tangential anisotropy. The action of tides results in more isotropic/tangential orbits in the outskirts of galaxies and lowers the size of the stellar population. We note that some of the ``true'' anisotropy profiles, with several inflection points, cannot be well described by the \citet{Baes_2007} parameterisation and would likely benefit from a more general approach (see section 2.1 of \citealt{Read_2017}). The anisotropy profiles in Draco, Ursa Minor and Sculptor are recovered well, reflecting the improved quality of the density profile recovery compared to Fornax and Carina.

\renewcommand\thefigure{C\arabic{figure}}

\section{Choice of Power law break radii}\label{sec:rbreak}

\begin{figure}
    \centering
    \includegraphics[width=\columnwidth]{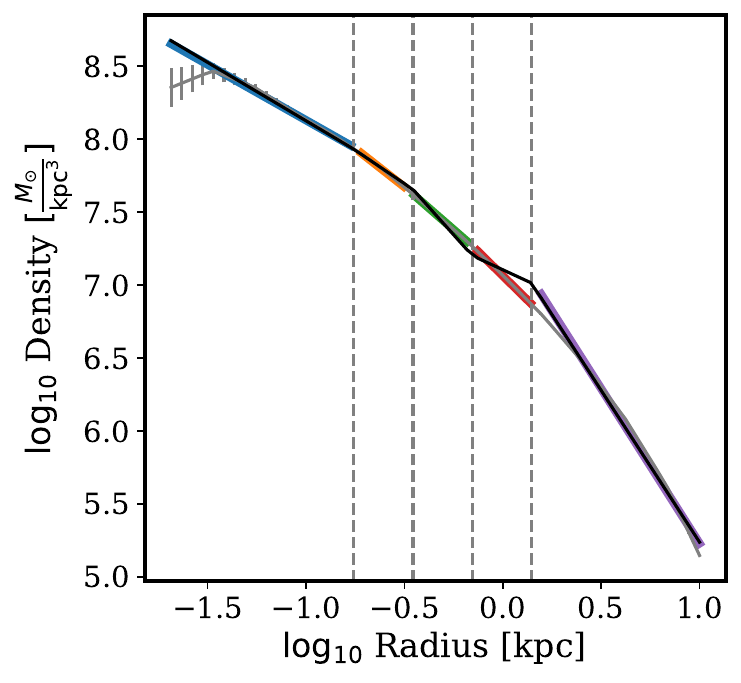}
    \caption{The density of the Fornax `light' model as a function of radius (grey line) fit by the broken-power-law model (black line). The grey dashed lines show the regions of the profile that are fit by each power-law slope $\gamma_i$ and correspond to multiples of $R_e$. Thick colour lines show the result of fitting a power law to each individual region of the profile.}
    \label{fig:powerlaws}
\end{figure}
\begin{figure}
    \centering
    \includegraphics[width=\columnwidth]{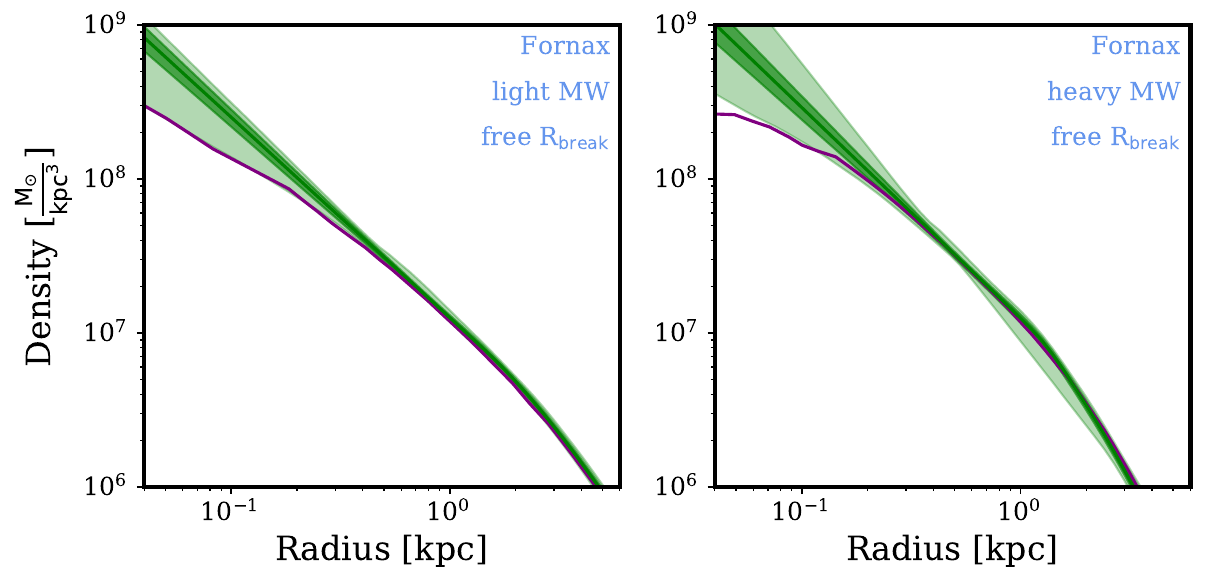}
    \caption{The result of applying a modification of {\sc pyGravSphere} to the Fornax `light' and `heavy' models, where $R_{\rm break}$ (which is typically set to the half-light radius) is allowed to be a free parameter of the model.}
    \label{fig:rbreak}
\end{figure}

In the original formulation of the {\sc GravSphere} model \citep{Read_2017}, the breaks in the broken-power-law density profile are defined as multiples of the projected half-light radius $[0.25,0.5,1,2,4]R_e$, motivated by the mass content being most strongly constrained within $R_e$. Potential dark matter core formation due to heating is also expected to occur within approximately the half-light radius of the dwarf, motivating additional binning in of the profile in this region. \citet{Read_2017} considered mock data with as much as $10^4$ stellar tracers and found that the choice of dark matter binning did not alter their results significantly. However, this result ultimately depends on how embedded the stellar profile is inside the dark matter halo and whether the multiples of $R_e$ can capture the major changes in the log-slope of the dark matter profile, particularly near the dark matter scale radius and the tidal radius. Compared to the mocks discussed in detail in \citet{Read_2017} and \citet{breaking_beta}, the half-light radii of our dwarfs are significantly more embedded within the dark matter scale radius (typically within 10~per~cent of the dark matter scale radius), meaning that 2-4$R_e$ does not capture the region beyond the dark matter scale radius. When the dwarfs are tidally stripped, the dark matter density profile is additionally steepened. The accuracy of the broken power-law fit therefore depends on whether the region 2-4$R_e$ can capture the significant change in the log-slope of the dark matter. In fact, a good predictor of {\sc GravSphere}'s performance on mocks analysed in \citet{breaking_beta} is how embedded the stellar half-light radius is within the dark matter profile scale radius, with the most inaccurate results arising from the cases where $r_s\gtrsim4R_e$. The availability of kinematic tracers at these radii likely has a secondary effect, whereby the virial shape parameters put a stronger constraint on the outer slope, which affects the recovery of the inner profile via the smoothness criterion.

In Fig.~\ref{fig:powerlaws} we investigate this by directly fitting the power law model to the known density profile of the Fornax `light' model at the present day, without the inclusion of the smoothness criterion. The `kick' in the profile near 1~kpc can be seen, which is more subtle in Fig.~\ref{fig:for_car_dens} (see also the Supplementary Materials). We then test this result by splitting the profile into regions that are fit by individual power laws. The first four power laws give an excellent fit to the profile, but the fifth power law ($\gamma_4$) attempts to fit the outermost steeper regions of the halo. It is straightforward to see how this results in the shape of the overall fit (black line) and how this would lead to inferred core-like profiles in the inner regions if slopes below $\gamma_3$ have to be shallower. In Supplementary Materials, we also show that applying the {\sc GravSphere} model without the smoothness criterion leads to a recovered profile similar to the black line in Fig.~\ref{fig:powerlaws}. 

This problem can of course be solved by adding additional bins to the power-law model. Ideally, a range of power-law bins can be tested to ensure convergence of the result. \citet{gravsphere2} have also utilised our Fornax `light' model to test the {\sc GravSphere2} code that utilises the {\sc coreNFWtides} dark matter model and achieved a more accurate dark matter profile recovery. Here, we also test a simple modification to the broken-power-law model, where the break radius $r_{\rm break}$ is treated as a free parameter, instead of being set to the projected half-light radius. As shown in Fig.~\ref{fig:rbreak}, this leads to an accurate recovery of the profile in the outer parts, while, at the centre, a more cuspy solution is preferred, since 0.25 times the best-fitting $r_{\rm break}$ corresponds to approximately 0.5~kpc and cannot fully capture the inner regions of the profile. 

This explains in part why the effect is less significant in Draco and Ursa Minor, since the data does not reach far enough to capture kinematic features associated with the change in dark matter density. As we show in the Supplementary Materials, data truncation to the central 2~kpc has resulted in more accurate recovery of the dark matter density in the Fornax `light' and `heavy' model. However, one has to bear in mind that the projected central regions of the dwarf also carry information about the outer density profile, and thus the bias cannot be avoided fully with data truncation when the model is a poor representation of the true dark matter density profile.



\FloatBarrier

\bsp	
\label{lastpage}
\end{document}